\documentclass[%
 reprint,
 amsmath,amssymb,
 aps,
]{revtex4-2}

\usepackage{float}
\usepackage{graphicx}
\usepackage{dcolumn}
\usepackage{bm}
\usepackage{amsmath, amssymb, physics, color, comment}
\usepackage{hyperref}

\usepackage{makecell}

\begin{document}
\preprint{APS/123-QED}

\title{Finite-energy Gottesman-Kitaev-Preskill state-enhanced optical interferometry}

\author{
 Ashmita Roy \\
  Inter-University Centre for Astronomy and Astrophysics (IUCAA),\\
 Post Bag 4, Ganeshkhind, Pune, 411007, India 
  \texttt{roy.ashmita@iucaa.in} \\
  R. Srikanth \\
  Theoretical Sciences Division, Poornaprajna Institute of Scientific Research\\
  Bidalur post, Devanahalli, Bengaluru 562164, India \\
  \texttt{srik@ppisr.res.in} \\
 Deepak Pandey \\
  Inter-University Centre for Astronomy and Astrophysics (IUCAA),\\
 Post Bag 4, Ganeshkhind, Pune, 411007, India 
  \texttt{deepak.pandey@iucaa.in} \\
  }


\date{\today}

\begin{abstract}
We present the case of a Gottesman-Kitaev-Preskill (GKP) state-enhanced optical interferometry with detailed analysis of the phase sensitivity for both the SU(2) and SU(1,1) interferometers. The conventional quantum-enhanced SU(2) interferometer, employing coherent light at one input port and squeezed light at the other, is compared with a modified configuration using coherent light and a GKP state. While it is known that the squeezed vacuum state is the optimal Gaussian resource input mode when paired with the coherent state, we show that the finite-energy GKP state with sufficiently broad envelope outperforms the squeezed vacuum injection, irrespective of the presence of optical losses. This can be attributed to the enhanced robustness coming from the availability of multiple squeezed peaks in the GKP case. However, because lowering the mean photon number reduces the GKP envelope width, the squeezed vacuum input performs better when compared with a GKP state of equal or lower mean photon number. We also observe that optical losses tend to diminish the relative advantage of either input state, since both states approach the (unsqueezed) vacuum state asymptotically. Our work demonstrates the direct application of finite-energy GKP states in optical interferometry along with a methodology for estimating the quantum Fisher information (QFI) and presents a phase estimation procedure using non-Gaussian resources in comparison with conventional Gaussian states. 

\end{abstract}

\maketitle


\section{\label{sec:level1}Introduction}
The precise estimation of physical parameters is a fundamental objective of both classical and quantum sensing \cite{Quantum_sensing_review}.  In optical metrology, the estimation of small phase shifts is of particular importance, as many physical quantities including displacement \cite{Displacement_sesning_review,Displacement_2}, acceleration \cite{modified_optical_accelerometer,Accelerometer_2}, rotation \cite{Gyroscope_review}, and gravitational-wave strain \cite{Freise2010} can be encoded in or inferred from changes in the optical phase. Interferometers provide a natural framework for such measurements by mapping the relative optical path-length difference between two modes onto a corresponding relative phase shift, which is subsequently encoded in the measurable interference signal at the output ports. As a result, interferometric phase estimation has become a cornerstone of optical sensing based precision measurements, driving advances in both fundamental physics and measurement technologies. 

Quantum noise arising from vacuum fluctuations imposes one of the fundamental sensitivity limits in precision optical interferometry \cite{Caves_1981}. Quantum-enhanced optical interferometry seeks to improve the measurement precision \cite{Giovannetti_2004_science} by exploiting non-classical resources, such as squeezing \cite{GW_Quantum_enhanced_PhysRevLett},  entanglement \cite{Lukin_2026} and non-Gaussian states \cite{Rana_2026}. For example, a widely adopted approach to mitigate  limitation in the phase measurement beyond the so called shot-noise limit, is the injection of a squeezed vacuum state into the second input port of the interferometer \cite{Caves_2013, GEO600_PhysRevA}. By suppressing quantum fluctuations in the quadrature relevant to the phase measurement below the vacuum-noise level, while increasing fluctuations in the conjugate quadrature, squeezed vacuum reduces the quantum noise contributing to the phase readout and can improve the shot-noise-limited sensitivity without increasing the circulating optical power \cite{Punturo_2010}. This technique has been successfully implemented in interferometers even for detection in a broad frequency band \cite{Broadband_squeezing_PhysRevX}. 

However, the performance of squeezed states is susceptible to optical loss \cite{Gao_2016, Hofmann_2010}. Loss introduces vacuum fluctuations \cite{Walmsley_PhysRevA} that degrade the squeezing, progressively moving the quadrature variance toward that of the ordinary vacuum state and thereby diminishing the achievable quantum enhancement. Consequently, understanding and mitigating the effects of optical loss remains a central challenge in the practical implementation of squeezed-light-enhanced interferometry \cite{Escher2011}. These limitations motivate the exploration of alternative non-classical resource states that exhibit greater robustness against optical loss while retaining the potential to enhance phase sensitivity. Owing to their inherent robustness against Gaussian noise and their demonstrated advantages in displacement sensing,  Gottesman-Kitaev-Preskill (GKP) states present an attractive alternative for precision metrology \cite{GKP_PRX_2026,Larsen2025}. Moreover, recent studies have shown that non-Gaussian states \cite{Displacement_nonGaussian, Non_Gaussian_Force_sensing} can achieve higher quantum Fisher information (QFI) than their Gaussian counterparts under comparable resource constraints, suggesting the possibility of enhanced phase sensitivity in practical interferometric settings. However, we find that the application of GKP states to interferometric phase estimation has received relatively less attention. 

In this work, we aim to determine the conditions under which and extent to which GKP states provide a metrological quantum advantage over the best Gaussian resource known in this context, namely the squeezed vacuum state. We derive analytical expressions for the QFI of both these states propagating through lossy SU(2) and SU(1,1) interferometers and compare their phase sensitivities under realistic conditions. We also show how the quantum advantage depends on the relative mean photon numbers of these two dark port input states and how channel loss can level out the quantum advantage.

\section{Preliminaries}
\subsection{Gottesman-Kitaev-Preskill States} \label{sec:prelim_gkp}
The Gottesman-Kitaev-Preskill (GKP) state was originally proposed as a bosonic quantum error-correcting code that encodes logical qubits into the continuous-variable quadratures of a harmonic oscillator \cite{GKP_2001}. Unlike Gaussian states, GKP states possess a non-Gaussian grid-like structure in phase space, consisting of a periodic array of sharply localized peaks modulated by a broad envelope. This unique structure endows them with remarkable robustness against small Gaussian displacement errors, making them a promising resource for fault-tolerant quantum information processing \cite{Sabapathy_PhysRevA, optical_GKP_2010}. More recently, GKP states have attracted significant interest in quantum metrology, where their resilience to Gaussian noise and exceptional displacement sensitivity have been shown to offer advantages in precision sensing \cite{GKP_PRX_2026, Mutliparameter_sensing}. The ideal canonical GKP state is invariant under discrete phase-space rotations, whose position-space wavefunction is given by
\begin{equation}
\braket{q}{\mathrm{GKP}_{\mathrm{ideal}}}
\propto
\sum_{s=-\infty}^{\infty}
\delta\left(q-s\sqrt{2\pi}\right).
\label{eq:qu0}
\end{equation}
The corresponding momentum-space wavefunction is obtained by Fourier transformation and has the reciprocal lattice structure, with identical spacing as above. 

In particular, the GKP state in EQ. (\ref{eq:qu0}) is the ``qunaught state'', characterized by a lattice cell of area $2\pi = \delta_q\cdot\delta_p$, where $\delta$ is the length of a single shift in $p$ or $q$ direction. Since both directions are symmetric, $\delta^2 = 2\pi$, giving us $\delta = \sqrt{2\pi}$ as the shift distance in the qunaught lattice. A \textit{computational} GKP lattice is a coarse-graining of the finer qunaught lattice, were each cell has an area $4\pi = \delta_q\cdot\delta_p = \delta^2$ , giving us $\delta=2\sqrt{\pi}$. This is crucial for the computational purpose because a shift of half that cell, i.e., $\sqrt{\pi}$, toggles the system between the logical-0 lattice and the logical-1 lattice. However, for our present interferometric (rather than computational) purpose, the qunaught GKP state is preferred.

The ideal state $\ket{\mathrm{GKP}_{\mathrm{ideal}}}$ in Eq. (\ref{eq:qu0}) is an unphysical state with infinite energy, lying outside the Hilbert space of normalizable states. In particular, $\braket{\mathrm{GKP}_{\mathrm{ideal}}}{\mathrm{GKP}_{\mathrm{ideal}}} = \infty$, and possesses infinite energy, $\braket{\hat{n}} \to \infty$. In practice, physically realizable approximate or finite-energy GKP states are obtained by damping the ideal state
by means of a non-unitary Gaussian damping operator in Fock space:
\begin{equation}
\ket{\mathrm{GKP}_{\Delta}}
=
\mathcal{N}_{\Delta}
e^{-\Delta^2 \hat{n}}
\ket{\mathrm{GKP}_{\mathrm{ideal}}},\label{eq:gkp_one_param}
\end{equation}
where $\hat{n}=\hat{a}^{\dagger}\hat{a}$ is the photon number operator, $\Delta>0$ is the regularization parameter, and $\mathcal{N}_{\Delta}$ is a normalization constant. The wavefunction of the finite-energy GKP state:
\begin{align}
\braket{q}{\mathrm{GKP}_{\Delta}}
&=
 \mathcal{N}_{\Delta} 
\exp\left(-\pi\Delta^2 q^2\right) \nonumber \\
&\times \sum_{s=-\infty}^{\infty}
\exp\left[
-\frac{\left(q-s\sqrt{2\pi}\right)^2}{2\Delta^2}
\right],
\label{eq:damp}
\end{align}
This state replaces the infinitely sharp lattice with finite-width peaks enclosed by a Gaussian envelope \cite{approx_gkp_2020, approx_gkp_2020_2}. The damped form Eq. (\ref{eq:damp}) is square-integrable, and hence normalizable with finite mean photon number. Here, we have
\begin{equation}
    \begin{aligned}
        \sigma_{\mathrm{peak}} = \Delta; \quad \sigma_{\mathrm{env}} = \frac{1}{\sqrt{2\pi}\Delta}.
    \end{aligned}
    \label{eq:peakenv}
\end{equation}
Accordingly, the width of the global envelope is inversely related to the width of the individual peaks such that their product is a constant: $\sigma_{\mathrm{peak}} \cdot \sigma_{\mathrm{env}} = \frac{1}{\sqrt{2\pi}}$. 

A squeezed finite-energy GKP state is obtained by applying a unitary squeezing operator $\hat{S}(\kappa)$ to the finite-energy GKP state:
\begin{equation}
\ket{\mathrm{GKP}_{\kappa,\Delta}}
=
\hat{S}(\kappa)
\ket{\mathrm{GKP}_{\Delta}}.
\end{equation}
For the momentum-squeezing convention used here, we have $\hat{S}^{\dagger}(\kappa)\hat{q}\hat{S}(\kappa) = e^{\kappa}\hat{q}$, and $\hat{S}^{\dagger}(\kappa)\hat{p}\hat{S}(\kappa) = e^{-\kappa}\hat{p}$. Such squeezed GKP states have been considered in both theoretical and experimental proposals for GKP-state generation and manipulation \cite{Winnel_PRL_2024, Kendal_APL_2024}. 

In the position basis, the squeezing operator acts as a scaling transformation. In place of Eq. (\ref{eq:damp}), we have:
\begin{align}
\braket{q}{\mathrm{GKP}_{\kappa,\Delta}}
&= \mathcal{N}_{\kappa,\Delta}
\exp\left[
-\frac{\pi\Delta^2q^2}{e^{2\kappa}}
\right] \nonumber \\
&\times \sum_{s=-\infty}^{\infty}
\exp\left[
-\frac{
\left(q-e^{\kappa}s\sqrt{2\pi}\right)^2
}{
2e^{2\kappa}\Delta^2
}
\right], \label{eq:gkp_2_param_wavefn}
\end{align}
where $\mathcal{N}_{\kappa,\Delta}$ is the normalization constant. Thus, squeezing scales the lattice spacing, the individual peak width, and the envelope width by the same factor $e^{\kappa}$. The distance between lattice sites is scaled as $d_{\mathrm{lattice}} = e^{\kappa}\sqrt{2\pi}$, the peak widths are scaled as
\begin{equation}
    \begin{aligned}
        \sigma_{\mathrm{peak}} = e^{\kappa}\Delta; \quad \sigma_{\mathrm{env}} = \frac{e^{\kappa}}{\sqrt{2\pi}\Delta}
    \end{aligned}
    \label{eq:peakenv+sq}
\end{equation}
in place of Eq. (\ref{eq:peakenv}). Thus squeezing changes the overall scale of the GKP lattice without changing its intrinsic comb structure.

In light of Eq. (\ref{eq:peakenv+sq}), the wavefunction (\ref{eq:gkp_2_param_wavefn}) can be rewritten as:
\begin{align}
\braket{q}{\mathrm{GKP}_{\sigma_{\mathrm{peak}},\sigma_{\mathrm{env}}}} &=
\mathcal{N}_{\sigma_{\mathrm{peak}},\sigma_{\mathrm{env}}} \exp\left( -\frac{q^2} {2\sigma_{\mathrm{env}}^2} \right)
\nonumber \\ &\times \sum_{s=-\infty}^{\infty}
\exp\left[
-\frac{\left(q-s\sqrt{2\pi}\right)^2}{2{\sigma_{\mathrm{peak}}^2}}
\right],\label{eq:gkp_two_width}
\end{align}
where $\sigma_{\mathrm{peak}}$, which controls the width of the individual peaks, and $\sigma_{\mathrm{env}}$ controls the extent of the global envelope, appear as the independent parameters instead of $\kappa, \Delta$.

\subsection{SU(2) and SU(1,1) Interferometers}\label{sec:SU_interferometers}

We consider two interferometer architectures for phase estimation. An SU(2) interferometer which is like a rotation on a sphere which splits and recombines light while preserving the total photon number over two modes. And an SU(1,1) interferometer which is like a hyperbolic rotation on a hyperboloid that amplifies and de-amplifies light while preserving the photon number difference. 

A beam splitter implements the SU(2) transformation
\begin{equation}
\begin{pmatrix}
a_{\mathrm{out}}\\
b_{\mathrm{out}}
\end{pmatrix}
=
\begin{pmatrix}
\cos\theta & \sin\theta\\
-\sin\theta & \cos\theta
\end{pmatrix}
\begin{pmatrix}
a_{\mathrm{in}}\\
b_{\mathrm{in}}
\end{pmatrix},
\label{eq:SU2_BS}
\end{equation}
where $\theta$ is the beam-splitter mixing angle. For a 50:50 beam splitter, $\theta=\pi/4$. The phase shift in one arm is described by $a\mapsto a e^{i\phi}$. The two beam splitters together with this phase shift form the conventional SU(2) interferometer.

In an SU(1,1) interferometer, the beam splitters are replaced by OPAs.
The corresponding Bogoliubov transformation can be written as
\begin{equation}
\begin{pmatrix}
a_{\mathrm{out}}\\
b_{\mathrm{out}}^\dagger
\end{pmatrix}
=
\begin{pmatrix}
\cosh r & e^{i\vartheta}\sinh r\\
e^{-i\vartheta}\sinh r & \cosh r
\end{pmatrix}
\begin{pmatrix}
a_{\mathrm{in}}\\
b_{\mathrm{in}}^\dagger
\end{pmatrix},
\label{eq:SU11_OPA}
\end{equation}
where $r$ is the parametric-amplification strength and $\vartheta$ is the pump phase. An SU(1,1) interferometer consists of two such OPAs separated by a phase shift, with the relative pump phase chosen such that the second OPA partially reverses the transformation of the first. The detailed input-output transformations are given in Appendices~\ref{sec:app_SU2} and~\ref{sec:app_SU11}.

This common form provides a convenient starting point for comparing different dark port input states. In particular, it allows the SU(2) and SU(1,1)
interferometers to be treated within the same phase-estimation
framework, as described below.
\subsection{Phase estimation and quantum Fisher Information}

In the small-phase and high coherent amplitude regime, both the SU(2) and SU(1,1) interferometers convert the unknown phase shift into a displacement of the dark-port
field. To first order in $\phi$, the output field can be written in the
common form
\begin{equation}
a_{\mathrm{out}}
\simeq
a_{\mathrm{in}}+iA\phi,
\label{eq:main_phase_encoding}
\end{equation}
where the phase-to-displacement coefficient is
\begin{equation}
A=
\begin{cases}
\dfrac{\alpha}{2}, & \mathrm{SU(2)},\\[2mm]
G\alpha, & \mathrm{SU(1,1)},
\end{cases}
\qquad
G=\sinh r\cosh r.
\label{eq:main_A_factor}
\end{equation}
Here, $\alpha$ is the coherent carrier amplitude and $r$ is the parametric-amplification strength of the SU(1,1) interferometer. Thus, both interferometer architectures have the same phase-encoding structure, with their difference captured by the coefficient $A$. Using the quadrature convention 
\begin{equation}
a=\frac{q+ip}{\sqrt{2}},
\end{equation}
the phase-induced field displacement in Eq.~\eqref{eq:main_phase_encoding}
corresponds to a displacement of the momentum quadrature,
\begin{equation}
\delta p=\sqrt{2}A\phi.
\end{equation}
The small-phase input-output relation in
Eq.~\eqref{eq:main_phase_encoding} provides a single-mode description of the phase encoding at the dark port. Within this description, the phase acts as a displacement of the momentum quadrature, generated by
\begin{equation}
H_{\mathrm{eff}}=\sqrt{2}Aq.
\label{eq:main_effective_generator}
\end{equation}

The quantum Fisher information (QFI) quantifies the ultimate
information that the dark port input state carries about the unknown phase. For a pure
input state undergoing unitary encoding generated by
$H_{\mathrm{eff}}$, the QFI is
\begin{equation}
F_Q^{(\phi)}
=
4\,\mathrm{Var}(H_{\mathrm{eff}}).
\label{eq:QFI_main}
\end{equation}
The corresponding quantum Cram\'er-Rao bound sets the ultimate
precision limit for unbiased phase estimation,
\begin{equation}
\Delta\phi_{\rm QCRB}
\geq
\frac{1}{\sqrt{\nu F_Q^{(\phi)}}},
\label{eq:QCRB_main}
\end{equation}
where $\nu$ is the number of independent repetitions. Thus, a larger
QFI corresponds to a smaller ultimate phase uncertainty.

Substituting the effective generator in Eq.~\eqref{eq:main_effective_generator}
into Eq.~\eqref{eq:QFI_main} gives the expression
\begin{equation}
F_Q^{(\phi)}
=
8A^2\,\mathrm{Var}(q).
\label{eq:QFI_main_effective}
\end{equation}
Hence, for a given interferometer and coherent carrier amplitude, the
phase-estimation performance is determined by the variance of the quadrature conjugate to the phase-induced displacement. This provides a common framework for comparing momentum-squeezed finite-energy GKP states and momentum-squeezed vacuum states.

\section{Analysis of GKP-state enhanced interferometers}


In this section, we evaluate the quantum Fisher information of momentum-squeezed finite-energy GKP states and compare it with that of momentum-squeezed vacuum states within the framework introduced in the previous section. The analysis is carried out for both SU(2) and SU(1,1) interferometers driven by coherent light, with the two architectures differing only through the phase-to-displacement coefficient while sharing the same QFI formalism.

The detailed QFI expressions for the two different input states, each considered separately at the dark port of the interferometer in the absence of optical loss, are derived in Appendix~\ref{sec:app_QFI_noloss}, while the corresponding expressions in the presence of optical loss are described in Appendix~\ref{sec:app_QFI_loss}. The resulting QFI expressions are summarized in Table~\ref{tab:qfi_summary}. To facilitate comparison at a fixed energy resource, the same results are also expressed in terms of the mean photon number in Table~\ref{tab:qfi_summary_photon}, with the corresponding photon-number relations derived in Appendix~\ref{app:QFI_meanphoton}.

These results provide a direct quantitative framework for assessing whether the non-Gaussian structure of finite-energy GKP states can offer a phase-sensing advantage over conventional squeezed-vacuum states at a fixed mean photon number.

\begin{table}[h]
\centering
\renewcommand{\arraystretch}{1.4}
\caption{Summary of the quantum Fisher information for momentum-squeezed finite-energy GKP states and squeezed vacuum states.}
\label{tab:qfi_summary}

\begin{tabular}{|c|c|c|}
\hline
\makecell[c]{
\textbf{Dark Port}\\ 
\textbf{Input State}}
&
\textbf{Lossless QFI}
& \makecell[c]{
\textbf{Approximate QFI}\\ 
\textbf{under optical loss}}
\\
\hline
\makecell[c]{Finite-energy \\GKP}
& \makecell[c]{$ 8A^2 {e^{2\kappa}} \mathrm{Var}(q)$\\
(Equation ~\eqref{eq:gkp_qfi_lossless})}
& \makecell[c]{$ 8A^2 \left[\eta e^{2\kappa} \mathrm{Var}(q) + \frac{1-\eta}{2}\right] $\\
(Equation ~\eqref{eq:loss_gkp_qfi})}
\\
\hline

\makecell[c]{Squeezed \\vacuum}
& \makecell[c]{$ 4A^2e^{2\kappa} $\\
(Equation ~\eqref{eq:sq_qfi_lossless})}
& \makecell{$ 4A^2 \left( \eta e^{2 \kappa} + 1-\eta \right)$\\
(Equation ~\eqref{eq:loss_sq_qfi})}
\\
\hline

\end{tabular}
\end{table}

\begin{table}[h]
\centering
\renewcommand{\arraystretch}{1.4}
\caption{Summary of the quantum Fisher information for momentum-squeezed finite-energy GKP states and squeezed vacuum states in terms of mean photon number.}
\label{tab:qfi_summary_photon}

\begin{tabular}{|c|c|c|}
\hline
\makecell[c]{
\textbf{Dark Port}\\ 
\textbf{Input State}}
&
\textbf{Lossless QFI}
& \makecell[c]{
\textbf{Approximate QFI}\\ \textbf{under optical loss}}
\\
\hline

\makecell[c]{Finite-energy \\ GKP}
& \makecell[c]{$16A^2 \frac{\bar n_{\rm GKP}+1/2}{1+e^{-4\kappa}}$\\
(Equation ~\eqref{eq:qfi_mean_photon})}
& \makecell[c]{$16A^2 \eta \frac{\bar n_{\rm GKP}+1/2}
{1+e^{-4\kappa}}$ \\$+ 4A^2(1-\eta)$\\
(Equation ~\eqref{eq:loss_qfi_mean_photon})}
\\
\hline

\makecell[c]{Squeezed \\vacuum}
& \makecell[c]{$4A^2 \left[ 2\bar n_{\rm sq} + 1 \right.$\\$ \left.+ 2\sqrt{\bar n_{\rm sq}(\bar n_{\rm sq}+1)}\right]$\\
(Equation ~\eqref{eq:sq_qfi_mean_photon})}
& \makecell{$ 4A^2 \left[ 1 + 2\eta \right.$\\
$ \left.\left(\bar n_{\rm sq} + \sqrt{ \bar n_{\rm sq} (\bar n_{\rm sq}+1)}\right)\right]$\\
(Equation ~\eqref{eq:sq_loss_qfi_mean_photon})}
\\
\hline

\end{tabular}
\end{table}

The parameter $A$ depends only on the interferometer configuration, as shown in \eqref{eq:main_A_factor} and is independent of the dark port input state. Consequently, all expressions in Tables ~\ref{tab:qfi_summary} and ~\ref{tab:qfi_summary_photon} apply directly to both interferometer configurations upon substituting the appropriate value of $A$.

We first examine the one-parameter finite-energy GKP state as shown in \eqref{eq:gkp_one_param}, where the peak and envelope widths are linked by a single squeezing parameter $\Delta$, and identify the conditions under which the GKP state outperforms the squeezed-vacuum reference. We then extend the analysis to the more general two-parameter GKP state as shown in \eqref{eq:gkp_two_width}, allowing the peak and envelope widths $\sigma_{\mathrm{peak}}$ and $\sigma_{\mathrm{env}}$ to vary independently in order to separate their individual contributions to the quantum Fisher information. Finally, we compare the two dark port input families at equal mean photon number to determine whether any observed enhancement persists under a fixed-resource criterion.


\subsection{Dependence of QFI on GKP parameter $\Delta$}


The
variance of the unsqueezed finite-energy GKP state is
\begin{align}
\mathrm{Var}(q)
&=
\frac{\Delta^2}{2D}
+
\frac{\pi}{2D^2} \times \nonumber \\
&\frac{
\displaystyle
\sum_{s,t=-\infty}^{\infty}
(s+t)^2
\exp\left[
-\frac{\pi(s-t)^2}{2\Delta^2}
-\frac{\pi^2\Delta^2(s+t)^2}{1+2\pi\Delta^4}
\right]
}{
\displaystyle
\sum_{s,t=-\infty}^{\infty}
\exp\left[
-\frac{\pi(s-t)^2}{2\Delta^2}
-\frac{\pi^2\Delta^2(s+t)^2}{1+2\pi\Delta^4}
\right]
} 
\label{eq:exact_variance_results}
\end{align}

The variance in Equation~(\ref{eq:exact_variance_results}) contains
an infinite double sum over the two lattice indices $s$ and $t$. A detailed derivation of the variance can be found in Appendix~\ref{app:gkp_variance} and the numerical approximation of the infinite sum has been discussed in Appendix \ref{app:numerical_approx}.

The momentum squeezing is then applied to the GKP state, which anti-squeezes the position quadrature relevant for the QFI. Thus, for a squeezing parameter $\kappa$, the position-quadrature variance is scaled as $\mathrm{Var}_{\kappa}(q) = e^{2\kappa}\mathrm{Var}(q)$. This squeezed variance is subsequently modified by optical loss. For
a loss channel with transmissivity $\eta$, the output quadrature
variance becomes
\begin{equation}
\mathrm{Var}_{\mathrm{GKP}}^{\mathrm{loss}}(q) = 
\eta e^{2\kappa}\mathrm{Var}(q)_{\mathrm{GKP}} +
\frac{1-\eta}{2},
\label{eq:gkp_loss_variance}
\end{equation}
and
\begin{equation}
\mathrm{Var}_{\mathrm{SQ}}^{\mathrm{loss}}(q) =
\frac{1}{2}
\left(\eta e^{2\kappa}+1-\eta \right),
\label{eq:sq_var_loss}
\end{equation}
respectively. The ratio of QFI of GKP state to that is squeezed vacuum state under optical loss is 
\begin{equation}
   \begin{aligned}
        \frac{F_{\mathrm{GKP}}}{F_{\mathrm{SQ}}} &= \frac{8A^2 \left(\eta  e^{2\kappa} \mathrm{Var}(q)_{\mathrm{GKP}} +\frac{(1-\eta)}{2}\right) }{ 4A^2 \left(\eta e^{2\kappa} +(1-\eta)\right)}\\
        \implies \frac{F_{\mathrm{GKP}}}{F_{\mathrm{SQ}}} &= \frac{2\eta  e^{2\kappa} \mathrm{Var}(q)_{\mathrm{GKP}} + {(1-\eta)} }{  \eta e^{2\kappa} +(1-\eta)}    \end{aligned}\label{eq:exact_qfi_ratio_numerical} 
\end{equation}

\begin{figure*}[ht]
    \centering
    \includegraphics[width=\linewidth]{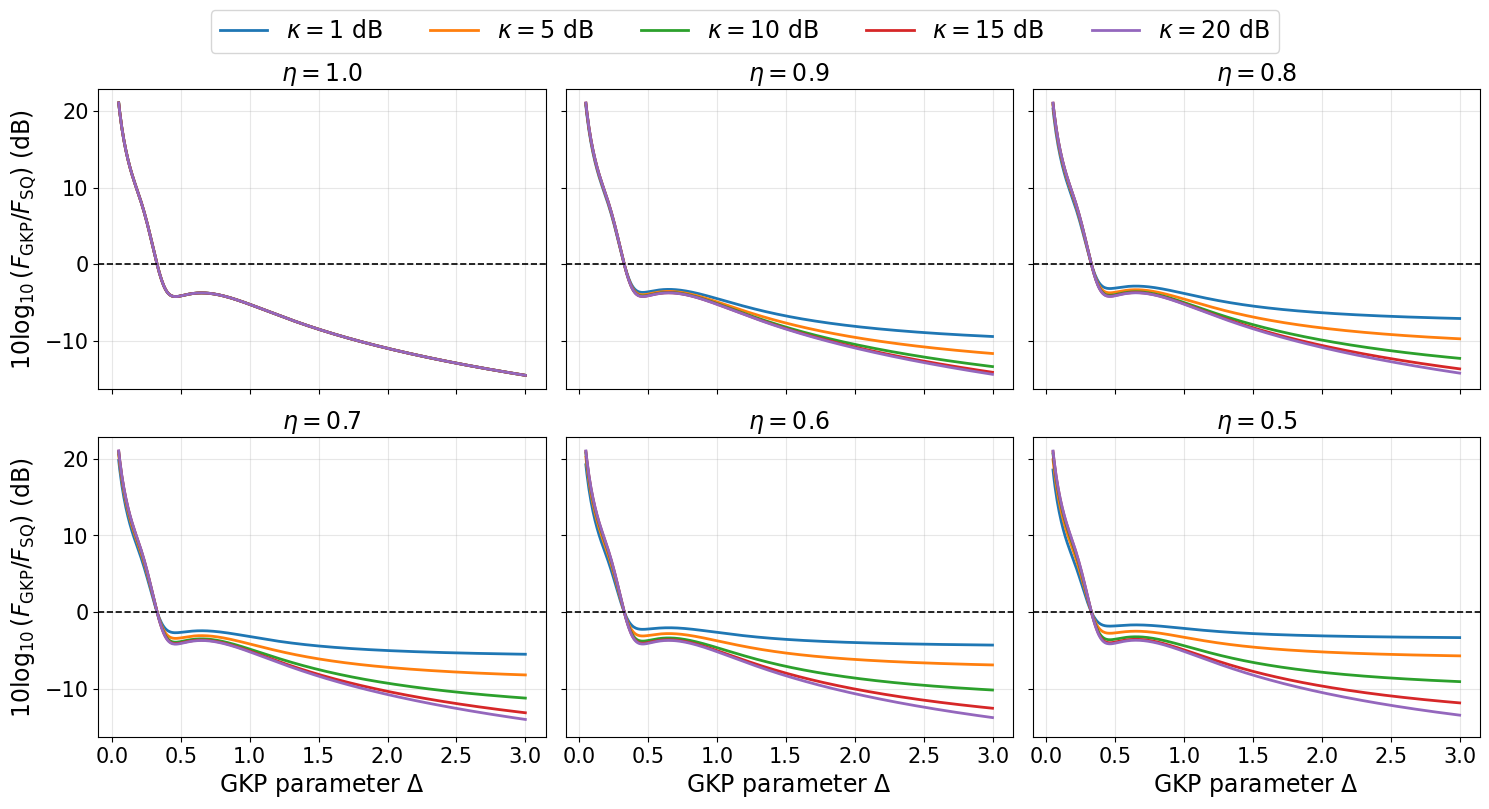}
    \caption{Quantum Fisher information ratio between the one-parameter finite-energy GKP state and the squeezed-vacuum state from Equation \eqref{eq:exact_qfi_ratio_numerical}, expressed in decibels as $10\log_{10} (F_Q^{\mathrm{GKP}}/F_Q^{\mathrm{sq}})$, as a function of the GKP damping parameter $\Delta$ for several values of the optical transmissivity $\eta$. Each panel corresponds to a different transmissivity, while the curves represent squeezing levels of $\kappa=1$, $5$, $10$, $15$, and $20~\mathrm{dB}$. The horizontal dashed line denotes equal QFI, corresponding to $F_Q^{\mathrm{GKP}}/F_Q^{\mathrm{sq}} = 1 $ (0 dB).}
    \label{fig:exact_qfi_ratio}
\end{figure*}

We plot the ratio of the QFI as a function of the GKP parameter $\Delta$ for several squeezing strengths and loss transmissivities in Figure ~\ref{fig:exact_qfi_ratio}. In the figure, the ratio has been converted to decibels as $10\log_{10}(F_{\mathrm{GKP}}/F_{\mathrm{SQ}})$, with $0$~dB corresponding to equal QFI. In the lossless case, $\eta=1$, all curves for different $\kappa$ essentially overlap.
This follows directly from Equation ~(\ref{eq:exact_qfi_ratio_numerical}),
since for $\eta=1$ the common factor $e^{2\kappa}$ cancels. Thus, in the absence of loss, squeezing changes the absolute QFI of both the dark port input states but does not change their relative sensitivity for a
fixed $\Delta$. 

For small $\Delta$, the QFI ratio is substantially above unity, indicating a strong advantage of the GKP state. As $\Delta$ is increased, the ratio decreases and crosses the equal-QFI threshold near $\Delta\simeq0.3$ - $0.4$. 

The crossover between the finite-energy GKP and squeezed-vacuum states can be understood directly in terms of their position-quadrature variances. The crossover occurs when the two dark port input states have equal QFI, i.e.,
\begin{equation}
\frac{F_{\mathrm{GKP}}}{F_{\mathrm{SQ}}}=1.
\end{equation}
Using Eq.~\eqref{eq:exact_qfi_ratio_numerical}, this condition becomes
\begin{align}
2\eta e^{2\kappa}
\mathrm{Var}_{\mathrm{GKP}}(q) + (1-\eta)
&=
\eta e^{2\kappa} + (1-\eta),
\end{align}
which gives
\begin{equation}
\mathrm{Var}_{\mathrm{GKP}}(q)=\frac{1}{2}.
\label{eq:equal_qfi_variance_condition}
\end{equation}
Thus, the GKP and squeezed-vacuum states have equal QFI precisely when the position-quadrature variance of the unsqueezed finite-energy GKP state equals the vacuum quadrature variance. Further, this crossover condition is independent of both the squeezing strength $\kappa$ and the loss transmissivity $\eta$, provided $\eta>0$.

Substituting the finite-energy GKP variance from Eq.~\eqref{eq:exact_variance_results} into Eq.~\eqref{eq:equal_qfi_variance_condition} and solving numerically gives
\begin{equation}
\Delta_{\mathrm{equal\;QFI}}\simeq0.3282.
\end{equation}
Consequently, 
\begin{equation}
F_{\mathrm{GKP}}>F_{\mathrm{SQ}}
\quad\Longleftrightarrow\quad
\mathrm{Var}_{\mathrm{GKP}}(q)>\frac{1}{2},
\end{equation}
which, for the finite-energy GKP states considered here, occurs for $\Delta<\Delta_{\mathrm{equal\;QFI}}\simeq0.3282$.

We now try to understand what is the physical meaning of this crossover in terms of the underlying GKP lattice structure. As discussed in Section~\ref{sec:prelim_gkp}, the lattice spacing is $d_{\mathrm{lattice}}=\sqrt{2\pi}$. Two distinct notions of localization are relevant here. The condition that individual lattice peaks be narrow compared with the lattice spacing is $\Delta\ll\sqrt{2\pi}$, whereas the condition that many lattice peaks lie within the overall probability envelope is $\Delta\ll\frac{1}{2\pi}$. The equal-QFI crossover therefore need not coincide with the regime in which the many-peak approximation is valid.

For the finite-energy GKP qunaught state, the probability density contains contributions from pairs of lattice components labelled by $s$ and $t$. The corresponding pair weight appearing in Eq.~\eqref{eq:exact_variance_results} is
\begin{equation}
W_{st} = \exp\left[ -\frac{\pi(s-t)^2}{2\Delta^2}
-\frac{\pi^2\Delta^2(s+t)^2}{1+2\pi\Delta^4} \right].
\end{equation}
The first factor,
\begin{equation}
\exp\left[ -\frac{\pi(s-t)^2}{2\Delta^2} \right],
\label{eq:overlap_suppression_factor}
\end{equation}
controls the suppression of contributions involving different lattice components. This dependence follows directly from the separation of the corresponding lattice centres $|q_s-q_t|=|s-t|\sqrt{2\pi}$,
so that $|s-t|$ measures the separation between the two components in units of the lattice spacing.

The diagonal terms, for which $s=t$, are therefore not suppressed by Eq.~\eqref{eq:overlap_suppression_factor}. For the off-diagonal terms, $s\neq t$, the largest contribution comes from nearest-neighbour components with $|s-t|=1$. Their overlap is suppressed by
\begin{equation}
\exp\left(-\frac{\pi}{2\Delta^2}\right),
\end{equation}
while more distant components are suppressed even more strongly. Hence, a sufficient condition for neglecting the off-diagonal contributions is
\begin{equation}
\exp\left(-\frac{\pi}{2\Delta^2}\right)\ll1
\quad\Longleftrightarrow\quad
\frac{\pi}{2\Delta^2}\gg1.
\label{eq:negligible_overlap_condition}
\end{equation}

At the equal-QFI crossover, $\Delta_{\mathrm{equal\;QFI}}\simeq0.3282$, the nearest-neighbour overlap factor is
\begin{equation}
\exp\left[
-\frac{\pi}{2(0.3282)^2}
\right]
\simeq4.6\times10^{-7}.
\end{equation}
Thus, the off-diagonal contributions are already extremely strongly suppressed at the crossover, and the individual lattice peaks can be regarded as effectively non-overlapping.

However, the number of lattice spacings contained within one standard deviation of the probability envelope at the crossover is
\begin{equation}
\frac{\sigma_{\mathrm{env}}}{d_{\mathrm{lattice}}}
=
\frac{1}{\Delta_{\mathrm{equal\;QFI}}}
\simeq3.04.
\end{equation}
This number is only a few lattice spacings and is therefore not parametrically large. Hence, although the individual GKP peaks are effectively non-overlapping at the equal-QFI crossover, the state does not yet lie in the asymptotic many-lattice-peaks regime.
\subsection{Effect of optical loss on the QFI comparison}
\label{sec:qfi_loss_effect}

\begin{figure*}[ht]
    \centering
    \includegraphics[width=\linewidth]{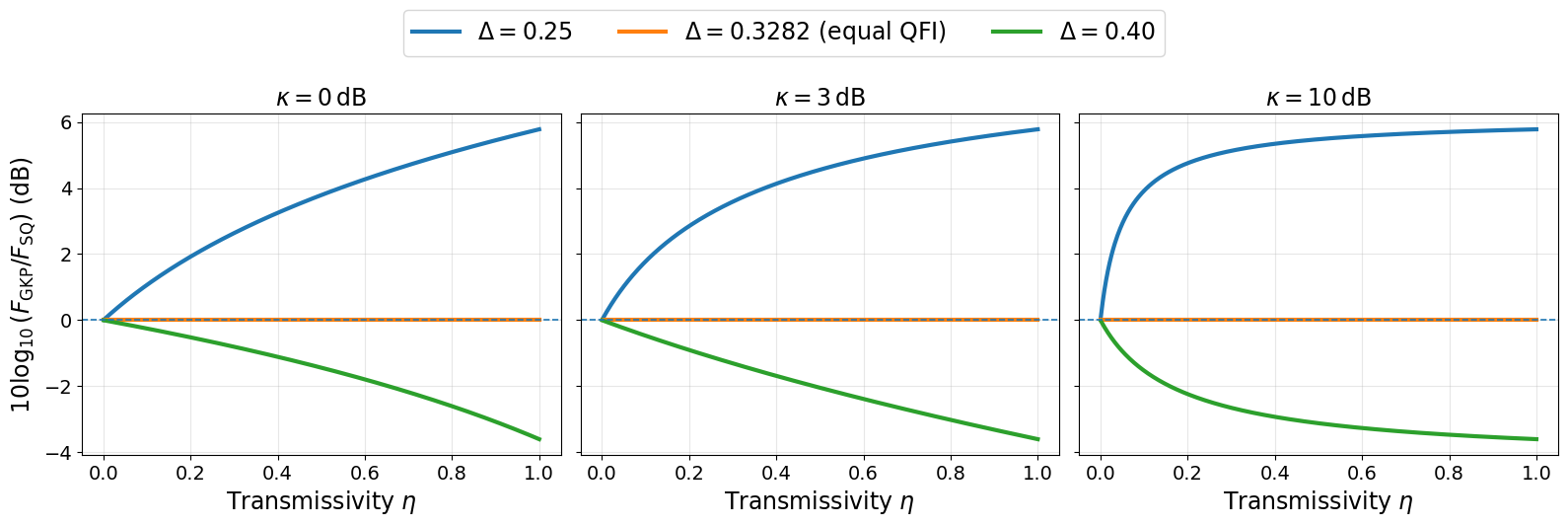}
    \caption{QFI ratio $F_{\rm GKP}/F_{\rm SQ}$, as in Equation \eqref{eq:exact_qfi_ratio_numerical}, expressed in dB, as a function of the optical transmissivity $\eta$ for three values of the GKP parameter: $\Delta=0.25$, $\Delta=\Delta_{\mathrm{equal\,QFI}}\simeq0.3282$, and $\Delta=0.40$. The three panels correspond to squeezing strengths $\kappa=0$, $3$, and $10$~dB, respectively. The dashed horizontal line denotes equal QFI, $F_{\rm GKP}/F_{\rm SQ}=1$. For     $\Delta<\Delta_{\mathrm{equal\,QFI}}$, the GKP state has a larger QFI, whereas for $\Delta>\Delta_{\mathrm{equal\,QFI}}$ the squeezed-vacuum state has a larger QFI. At the crossover   $\Delta_{\mathrm{equal\,QFI}}$, the two states have equal QFI for all $\eta$ and $\kappa$. Increasing optical loss suppresses the QFI difference between the two states, with $F_{\rm GKP}/F_{\rm SQ}\rightarrow1$ in the complete-loss limit $\eta\rightarrow0$}
\label{fig:qfi_loss_crossover}
\end{figure*}

Having established the equal-QFI crossover in terms of the intrinsic
position-quadrature variance of the finite-energy GKP state, we now
examine how optical loss affects the relative QFI of the two states.
The loss dependence is already contained in
Eq.~\eqref{eq:exact_qfi_ratio_numerical}. We can rewrite the expression as
\begin{equation}
\frac{F_{\mathrm{GKP}}}{F_{\mathrm{SQ}}}-1
=
\frac{
\eta e^{2\kappa}
\left[
2\mathrm{Var}_{\mathrm{GKP}}(q)-1
\right]
}{
\eta e^{2\kappa}+1-\eta
}.
\label{eq:qfi_ratio_loss_difference}
\end{equation}
This form makes the effect of loss particularly transparent. The
quantity determining the sign of the QFI difference is
$2\mathrm{Var}_{\mathrm{GKP}}(q)-1$, while the transmissivity $\eta$
controls the magnitude of the difference. Thus, optical loss does not
shift the equal-QFI crossover in $\Delta$ but rather, it progressively
reduces the relative difference between the two states.

Figure~\ref{fig:qfi_loss_crossover} illustrates this behavior by
plotting $10\log_{10}(F_{\mathrm{GKP}}/F_{\mathrm{SQ}})$ as a function
of the transmissivity $\eta$ for three representative values of
$\Delta$, chosen below, at, and above the equal-QFI crossover. The
three panels correspond to $\kappa=0$, $3$, and $10$~dB. The dashed
horizontal line denotes equal QFI.

For $\Delta=0.25$, which lies below the crossover
$\Delta_{\mathrm{equal\,QFI}}\simeq0.3282$, the unsqueezed GKP state
has $\mathrm{Var}_{\mathrm{GKP}}(q)>1/2$. Equation
\eqref{eq:qfi_ratio_loss_difference} therefore gives
$F_{\mathrm{GKP}}>F_{\mathrm{SQ}}$, and the corresponding curve lies
above the equal-QFI line. In contrast, for $\Delta=0.40$, the GKP
variance is smaller than the vacuum variance,
$\mathrm{Var}_{\mathrm{GKP}}(q)<1/2$, giving
$F_{\mathrm{GKP}}<F_{\mathrm{SQ}}$. 

The middle curve corresponds to the exact equal-QFI value $\Delta_{\mathrm{equal\,QFI}}\simeq0.3282$ for which
$\mathrm{Var}_{\mathrm{GKP}}(q)=1/2$. Substitution into
Eq.~\eqref{eq:qfi_ratio_loss_difference} gives
\begin{equation}
\frac{F_{\mathrm{GKP}}}{F_{\mathrm{SQ}}}-1=0,
\end{equation}
independently of both $\eta$ and $\kappa$. The equal-QFI crossover is
therefore unchanged by optical loss within the present
effective-displacement description. 

The reduction of the QFI difference with increasing loss can be
understood directly from the loss transformation of the quadrature
variance. For the GKP state,
\begin{equation}
\mathrm{Var}_{\mathrm{GKP}}^{\eta}(q)
-\frac{1}{2}
=
\eta
\left[
e^{2\kappa}\mathrm{Var}_{\mathrm{GKP}}(q)
-\frac{1}{2}
\right].
\label{eq:gkp_variance_loss_from_vacuum}
\end{equation}
Thus, loss continuously pulls the output variance toward the vacuum
variance $1/2$. The same process occurs for squeezed-vacuum. Consequently, for $\eta>0$, the relative difference between the QFI of finite-energy GKP and squeezed vacuum becomes smaller as
the transmissivity decreases. Thus,
\begin{equation}
{
\lim_{\eta\rightarrow0}
\frac{F_{\mathrm{GKP}}}{F_{\mathrm{SQ}}}=1.
}
\label{eq:qfi_ratio_complete_loss}
\end{equation}
Therefore, complete optical loss removes any relative QFI advantage:
the output states are dominated by the vacuum contribution introduced
by the loss channel, irrespective of the initial dark port input state.

The role of squeezing can also be seen in
Fig.~\ref{fig:qfi_loss_crossover}. For $\eta<1$, the vacuum contribution introduced by loss competes with the amplified quadrature variance. Increasing $\kappa$ therefore
increases the separation between the GKP and squeezed-vacuum QFIs at a
given nonzero loss, as seen from the progressively larger separation
of the curves in the higher-squeezing panels.

Overall, optical loss does not alter the intrinsic GKP parameter at
which the two dark port input states have equal QFI. Instead, it suppresses the relative QFI difference by replacing part of the dark port input state fluctuations with vacuum fluctuations. The GKP advantage for $\Delta<\Delta_{\mathrm{equal\,QFI}}$ and the squeezed-vacuum advantage for $\Delta>\Delta_{\mathrm{equal\,QFI}}$ therefore persist for nonzero transmissivity, while both advantages vanish in
the complete-loss limit.

\subsection{QFI dependence on independent peak and envelope widths}
\label{sec:gkp_two_width_results}

The generalized finite-energy GKP state introduced in Eq.~\eqref{eq:gkp_two_width} allows the peak width $\sigma_{\mathrm{peak}}$ and the envelope width $\sigma_{\mathrm{env}}$ to be varied independently.  This provides a two-parameter description in which the separate roles of the individual peak width and the overall lattice-envelope width can be examined. Since the QFI in the present effective-displacement framework is determined by the position-quadrature variance, this also allows us to identify how each width affects the phase-sensing performance.

The position variance for the two-parameter GKP state, derived in Appendix~\ref{app:gkp_variance_two_parameters}, is \begin{align}
&\mathrm{Var}_{\mathrm{GKP}}(q) = \frac{
\sigma_{\mathrm{peak}}^2 \sigma_{\mathrm{env}}^2 }{
2\left( \sigma_{\mathrm{peak}}^2+ \sigma_{\mathrm{env}}^2 \right) }
+ \frac{ \pi\sigma_{\mathrm{env}}^4 }{ 2\left(
\sigma_{\mathrm{peak}}^2+ \sigma_{\mathrm{env}}^2 \right)^2 } \times \nonumber \\
&\frac{ \displaystyle \sum_{s,t}
(s+t)^2 \exp\left[ -\frac{\pi(s-t)^2} {2\sigma_{\mathrm{peak}}^2} -\frac{\pi(s+t)^2}
{2\left( \sigma_{\mathrm{peak}}^2+  \sigma_{\mathrm{env}}^2 \right)} \right] }{
\displaystyle \sum_{s,t} \exp\left[ -\frac{\pi(s-t)^2}
{2\sigma_{\mathrm{peak}}^2} -\frac{\pi(s+t)^2}
{2\left( \sigma_{\mathrm{peak}}^2+ \sigma_{\mathrm{env}}^2
\right)} \right] }.
\label{eq:gkp_two_width_variance_results}
\end{align}

For the interferometric configuration considered here, the QFI is proportional to this position variance. After including squeezing and optical loss, the QFI relative to the squeezed-vacuum reference is
\begin{equation}
\frac{F_{\mathrm{GKP}}}{F_{\mathrm{SQ}}} = \frac{
2\eta e^{2\kappa}\mathrm{Var}_{\mathrm{GKP}}(q) +
(1-\eta) }{ \eta e^{2\kappa} + (1-\eta) }.
\label{eq:qfi_ratio_two_widths}
\end{equation}
Thus, for fixed squeezing $\kappa$ and transmissivity $\eta$, the dependence of the QFI ratio on  $\sigma_{\mathrm{peak}}$ and $\sigma_{\mathrm{env}}$ is entirely determined by their effect on the GKP position variance.

\begin{figure*}[t]
    \centering

    \includegraphics[width=0.49\textwidth]{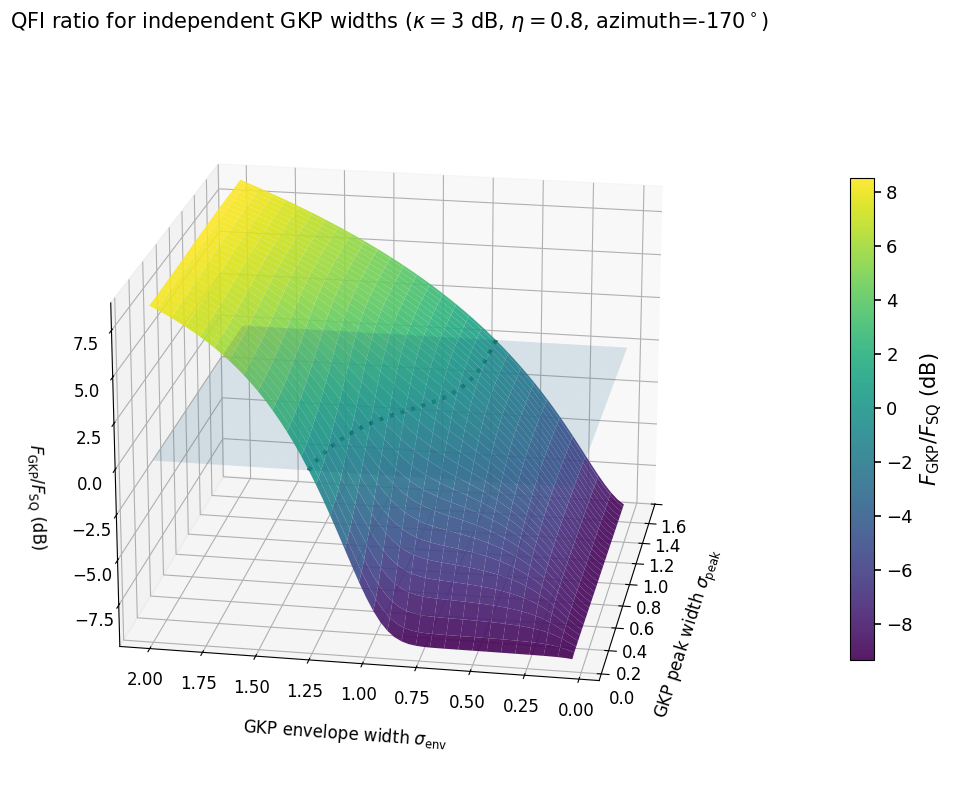}
    \hfill
    \includegraphics[width=0.49\textwidth]{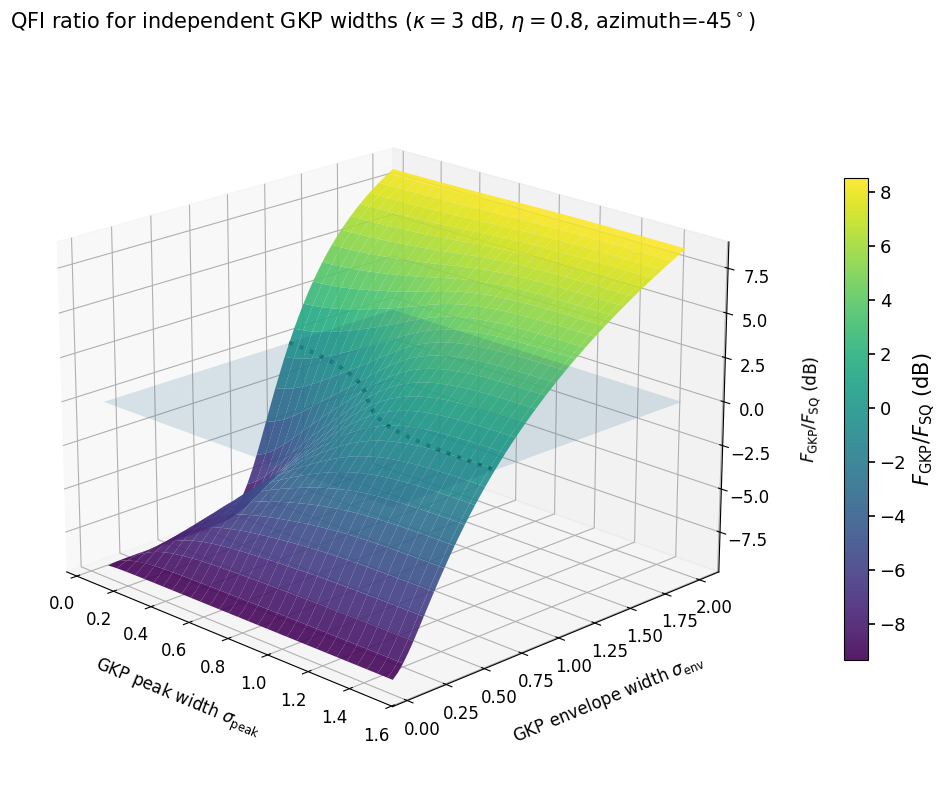}

    \vspace{0.5em}

    \includegraphics[width=0.49\textwidth]{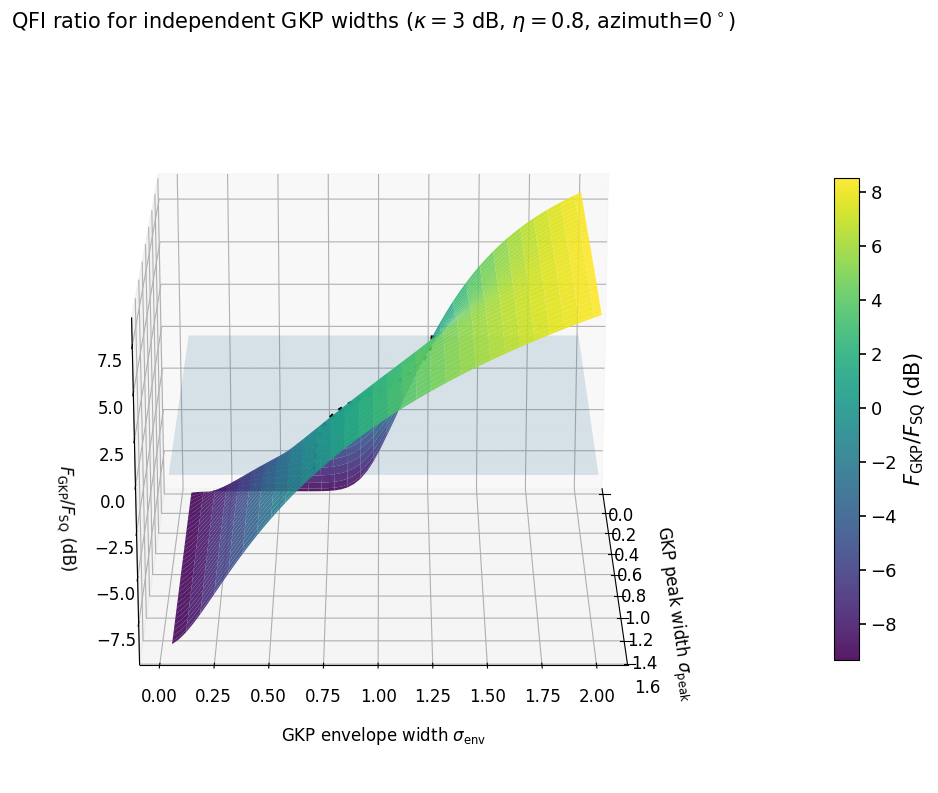}
    \hfill
    \includegraphics[width=0.49\textwidth]{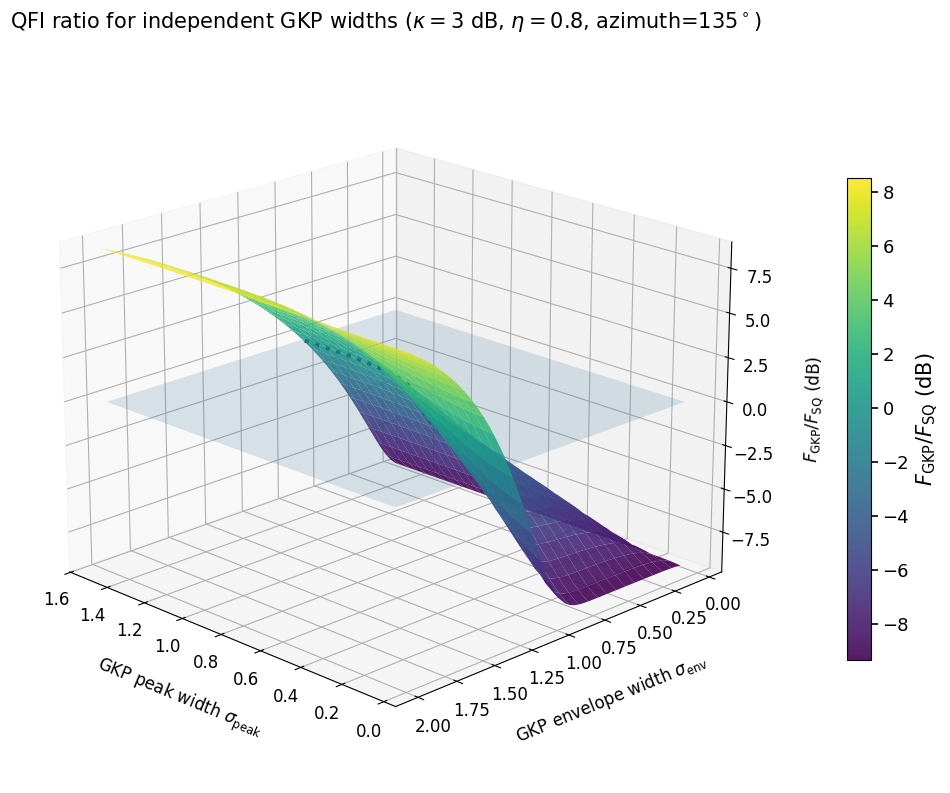}

    \caption{
        Three-dimensional representation of the QFI ratio
        $F_{\rm GKP}/F_{\rm SQ}$ as a function of the independent
        GKP peak and envelope widths, $\sigma_{\rm peak}$ and
        $\sigma_{\rm env}$, respectively. The squeezing is fixed at
        $\kappa=3$~dB and the transmissivity at $\eta=0.8$.
        The four panels show different azimuthal viewing angles of
        the same QFI surface, namely
        $-170^\circ$, $-45^\circ$, $0^\circ$, and $135^\circ$,
        respectively. The horizontal plane corresponds to
        $F_{\rm GKP}/F_{\rm SQ}=1$ (0~dB), while the dotted curve
        marks the equal-QFI boundary.
    }
    \label{fig:qfi_3d_rotation}
\end{figure*}

Figure~\ref{fig:qfi_3d_rotation} shows $10\log_{10}(F_{\mathrm{GKP}}/F_{\mathrm{SQ}})$ as a function of the two widths. The $0$-dB plane corresponds to equal QFI, $F_{\mathrm{GKP}}/F_{\mathrm{SQ}}=1$. Regions above this plane correspond to a larger QFI for the GKP state, while regions below it correspond to a larger QFI for the squeezed-vacuum state. The most pronounced dependence is on the envelope width $\sigma_{\mathrm{env}}$. Increasing the envelope width generally raises the QFI ratio. The origin of this behavior can be seen directly from
Eq.~\eqref{eq:gkp_two_width_variance_results}. The first term, $(\sigma_{\mathrm{peak}}^2 \sigma_{\mathrm{env}}^2)/{ 2( \sigma_{\mathrm{peak}}^2+ \sigma_{\mathrm{env}}^2 ) }$
describes the local contribution associated with the finite widths of the lattice components. For fixed  $\sigma_{\mathrm{peak}}$, this contribution increases with $\sigma_{\mathrm{env}}$ and approaches $\sigma_{\mathrm{peak}}^2/2$ when $\sigma_{\mathrm{env}}\gg\sigma_{\mathrm{peak}}$.

The second term captures the contribution from the distribution of the lattice components themselves. It contains the prefactor 
\begin{equation*}
\frac{ \pi\sigma_{\mathrm{env}}^4  }{
2\left( \sigma_{\mathrm{peak}}^2+ \sigma_{\mathrm{env}}^2
\right)^2 }
\end{equation*}
and the weighted lattice moment
\begin{align*}
&\frac{ \displaystyle \sum_{s,t}(s+t)^2 \exp\left[ -\frac{\pi(s-t)^2}  {2\sigma_{\mathrm{peak}}^2} -\frac{\pi(s+t)^2} {2\left(
\sigma_{\mathrm{peak}}^2+ \sigma_{\mathrm{env}}^2 \right)}
\right] }{
\displaystyle \sum_{s,t}\exp\left[ -\frac{\pi(s-t)^2}  {2\sigma_{\mathrm{peak}}^2} -\frac{\pi(s+t)^2} {2\left(
\sigma_{\mathrm{peak}}^2+ \sigma_{\mathrm{env}}^2 \right)}
\right] }.
\end{align*}
Increasing $\sigma_{\mathrm{env}}$ weakens the suppression of lattice components with large values of $|s+t|$. More components can therefore contribute to the overall lattice distribution, increasing its spatial extent and hence the position variance. Thus, the increase of the QFI with envelope width is not simply due to broader individual peaks, it also reflects the increased spatial extent of the populated GKP lattice.

The effect of $\sigma_{\mathrm{peak}}$ is different. Increasing $\sigma_{\mathrm{peak}}$ directly broadens the individual lattice peaks, modifying the local contribution to the variance. At the same time, it changes the overlap between neighboring lattice components through the factor $\exp[ -{\pi(s-t)^2}/ {2\sigma_{\mathrm{peak}}^2} ].$

For small $\sigma_{\mathrm{peak}}$, this factor strongly suppresses terms with $s\neq t$, corresponding to well-separated lattice peaks. As $\sigma_{\mathrm{peak}}$ increases, this suppression becomes weaker and neighboring peaks overlap more strongly. The dependence of the QFI on $\sigma_{\mathrm{peak}}$ therefore reflects two competing
effects: the direct broadening of individual peaks and the increasing overlap between different lattice components.

Because both widths influence the variance, equal QFI is not obtained at a single value of either parameter. Instead, it defines a boundary in the two-dimensional  parameter space,
\begin{equation}
\frac{ F_{\mathrm{GKP}} (\sigma_{\mathrm{peak}},\sigma_{\mathrm{env}}) }{ F_{\mathrm{SQ}} } =1. \label{eq:equal_qfi_boundary}
\end{equation}
This boundary is represented by the intersection of the QFI surface with the $0$-dB plane in  Fig.~\ref{fig:qfi_3d_rotation}. 

The connection with the original one-parameter finite-energy GKP state
is obtained by substituting $\sigma_{\mathrm{peak}}=\Delta$ and $\sigma_{\mathrm{env}} = {1}/{(\sqrt{2\pi}\Delta)}$ so that $\sigma_{\mathrm{peak}}\sigma_{\mathrm{env}} =
{1}/{\sqrt{2\pi}}$. The original GKP family therefore occupies only a one-dimensional curve in the two-dimensional $(\sigma_{\mathrm{peak}}, \sigma_{\mathrm{env}})$ parameter space explored in
Fig.~\ref{fig:qfi_3d_rotation}. 

Finally, the comparison in this section is performed at fixed $\kappa$ and $\eta$, rather than at fixed mean photon number. Since $\sigma_{\mathrm{peak}}$ and  $\sigma_{\mathrm{env}}$ are varied independently, changing either parameter changes the energy of the GKP state. Therefore, a region with $F_{\mathrm{GKP}}>F_{\mathrm{SQ}}$ only shows that the GKP state has a larger QFI for the specified state parameters; it does not by itself demonstrate an advantage at equal mean photon number. A resource-matched comparison requires us to compare the QFI at fixed the mean photon numbers.

\subsection{Dependence of QFI on Mean Photon Number}

\begin{figure*}[ht]
    \centering
    \includegraphics[width=\linewidth]{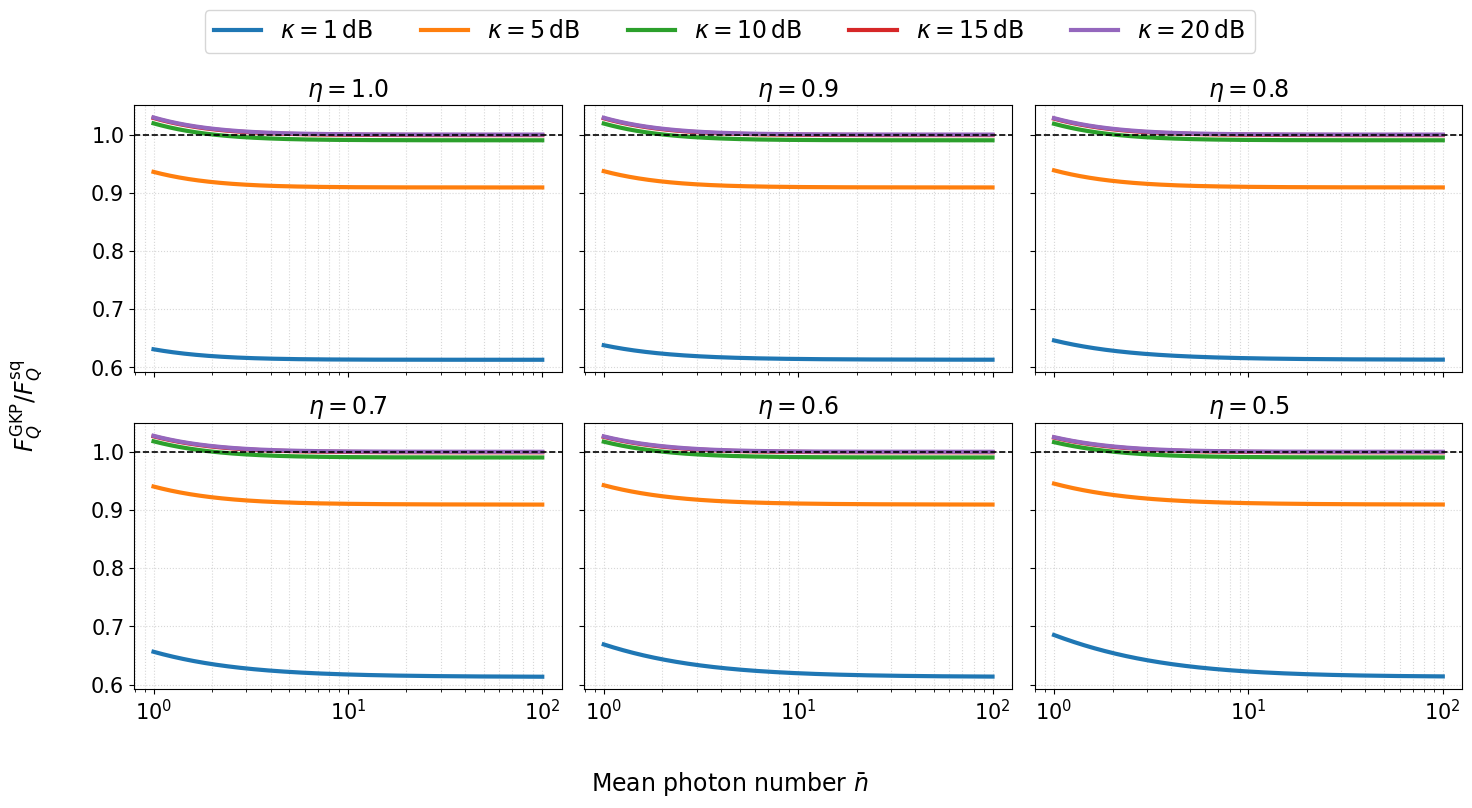}
    \caption{Ratio of the quantum Fisher information of the one-parameter finite-energy GKP state to that of a squeezed-vacuum state at equal mean photon number (Equation \eqref{eq:gkp_sq_qfi_ratio}), plotted as a function of $\bar n$ for several values of the momentum-squeezing parameter $\kappa$. The dashed line corresponds to equal QFI, $F_Q^{\mathrm{GKP}}=F_Q^{\mathrm{sq}}$. Each panel corresponds to different transmissivity $\eta$.}
    \label{fig:nbar_one_param}
\end{figure*}

The comparison at fixed mean photon number provides a more direct measure of the metrological performance of the GKP and squeezed-vacuum states. Detailed derivations can be found in Appendix \ref{app:QFI_meanphoton}. For the one-parameter finite-energy GKP state shown in \eqref{eq:gkp_one_param}, the ratio of the approximate QFI under optical loss to that of a squeezed-vacuum state 
with the same mean photon number (when $\bar n_{\mathrm{GKP}} = \bar n_{\mathrm{sq}}$) is
\begin{align}
    \frac{F_Q^{\mathrm{GKP}}}{F_Q^{\mathrm{sq}}} &=
    \frac{ 4\eta\left(\bar n_{\mathrm{GKP}}+\frac{1}{2}\right)+(1-\eta) }{
    \left(1+e^{-4\kappa_{\mathrm{GKP}}}\right) \left[ 1+2\eta\left(
    \bar n_{\mathrm{sq}}+\sqrt{\bar n_{\mathrm{sq}}(\bar n_{\mathrm{sq}}+1)} \right) \right] } \nonumber\\
    & = \frac{ 4\eta\left(\bar n+\frac{1}{2}\right)+(1-\eta) }{
    \left(1+e^{-4\kappa_{\mathrm{GKP}}}\right) \left[ 1+2\eta\left(
    \bar n+\sqrt{\bar n(\bar n+1)} \right) \right] }.
    \label{eq:gkp_sq_qfi_ratio}
\end{align}

For large mean photon number,
\begin{equation} 
\sqrt{\bar n(\bar n+1)} = \bar n\sqrt{1+\frac{1}{\bar n}}
\simeq \bar n+\frac{1}{2}, \qquad \bar n\gg1.
\end{equation}
Therefore, for any fixed nonzero transmissivity $\eta$,
\begin{equation}
\lim_{\bar n\rightarrow\infty} \frac{F_Q^{\mathrm{GKP}}}{F_Q^{\mathrm{sq}}} = \frac{1}{1+e^{-4\kappa_{\mathrm{GKP}}}}.
\label{eq:gkp_sq_qfi_ratio_asymptotic}
\end{equation}
In particular, this ratio approaches $1/2$ in the absence of
squeezing and approaches unity for strong squeezing,
\begin{equation}
\lim_{\kappa\rightarrow\infty} 
\lim_{\bar n\rightarrow\infty}
\frac{F_Q^{\mathrm{GKP}}}{F_Q^{\mathrm{sq}}} = 1.
\end{equation}
Thus, increasing the
squeezing allows the GKP state to approach the squeezed-vacuum QFI at
large photon number, but does not produce a parametrically larger QFI
in the asymptotic regime.

A more general conclusion can be obtained from the two-width
description as in \eqref{eq:gkp_two_width}, in which $\sigma_{\mathrm{peak}}$ and
$\sigma_{\mathrm{env}}$ are treated as independent parameters. More details can be found in Appendix \ref{app:GKP_mean_photon_two_widths}. Let
$\mathrm{Var}(q)$ and $\mathrm{Var}(p)$ denote the quadrature variances before
momentum squeezing. After squeezing,
\begin{equation}
\mathrm{Var}_{\kappa}(q)=e^{2\kappa}\mathrm{Var}(q),
\qquad
\mathrm{Var}_{\kappa}(p)=e^{-2\kappa}\mathrm{Var}(p).
\end{equation}
The mean photon number and the lossless displacement-sensing QFI are
then
\begin{equation}
\bar n_{\mathrm{GKP}}
=
\frac{\mathrm{Var}_{\kappa}(q)+\mathrm{Var}_{\kappa}(p)-1}{2},
\;
F_Q^{\mathrm{GKP}}
=
8A^2\mathrm{Var}_{\kappa}(q).
\end{equation}
Thus, at fixed mean photon number,
\begin{equation}
\mathrm{Var}_{\kappa}(q)+\mathrm{Var}_{\kappa}(p)=2\bar n_{\mathrm{GKP}}+1.
\label{eq:fixed_energy_variance_sum}
\end{equation}

By the Heisenberg uncertainty principle, the quadrature variances obey the relation
\begin{equation}
\mathrm{Var}_{\kappa}(q) \mathrm{Var}_{\kappa}(p)\geq\frac{1}{4}.
\end{equation}
Using Eq.~\eqref{eq:fixed_energy_variance_sum} to eliminate $\mathrm{Var}_{\kappa}(p)$ gives
\begin{equation}
\mathrm{Var}_{\kappa}(q)
\left(
2\bar n_{\mathrm{GKP}}+1-\mathrm{Var}_{\kappa}(q)
\right)
\geq
\frac{1}{4}.
\end{equation}
Solving this inequality for $\mathrm{Var}_{\kappa}(q)$ gives the upper bound
\begin{equation}
\mathrm{Var}_{\kappa}(q) \leq \bar n_{\mathrm{GKP}} +\frac{1}{2}
+ \sqrt{ \bar n_{\mathrm{GKP}} (\bar n_{\mathrm{GKP}}+1)
}. \label{eq:gkp_max_vq_fixed_energy}
\end{equation}
Since the QFI is proportional to $\mathrm{Var}_{\kappa}(q)$, this immediately gives
\begin{align*}
F_Q^{\mathrm{GKP}} &\leq 8A^2 \left[
\bar n_{\mathrm{GKP}} +\frac{1}{2} + \sqrt{
\bar n_{\mathrm{GKP}} (\bar n_{\mathrm{GKP}}+1) } \right]\\
&= 4A^2 \left[ 2\bar n_{\mathrm{GKP}}+1
+ 2\sqrt{ \bar n_{\mathrm{GKP}} (\bar n_{\mathrm{GKP}}+1) }
\right]
\end{align*}

The right-hand side is exactly the lossless QFI of a squeezed-vacuum
state with the same mean photon number,
\begin{equation*}
F_Q^{\mathrm{sq}} =
4A^2 \left[ 2\bar n_{\mathrm{sq}}+1 +
2\sqrt{ \bar n_{\mathrm{sq}} (\bar n_{\mathrm{sq}}+1) } \right].
\end{equation*}
Hence, at equal mean photon number,
\begin{equation}
F_Q^{\mathrm{GKP}} \leq F_Q^{\mathrm{sq}}.
\label{eq:gkp_sq_fixed_energy_bound}
\end{equation}

Equality requires $\mathrm{Var}_{\kappa}(q)\mathrm{Var}_{\kappa}(p)={1}/{4}$ that is, saturation of the minimum-uncertainty relation. A
squeezed-vacuum state is a pure Gaussian minimum-uncertainty state and therefore saturates this bound. Thus, within the displacement-sensing configuration considered here, the independently parameterized two-width GKP state cannot exceed the lossless QFI of a squeezed-vacuum state at fixed mean photon number.

This bound also clarifies the behavior observed numerically in Figure \ref{fig:nbar_one_param}. Although the GKP state can exhibit a large QFI for suitable choices of its peak and envelope widths, increasing the available photon number does not allow it to surpass the squeezed-vacuum benchmark when the comparison
is made at equal energy. The relevant advantage of the GKP structure must therefore be sought in other regimes or resource constraints, rather than at fixed mean photon number.

\section{Conclusions}

We have investigated  quantum-Fisher-information (QFI) improvement for phase estimation when finite-energy Gottesman-Kitaev-Preskill (GKP) states, rather than squeezed-vacuum states, are injected into the dark-port input of an interferometer driven by coherent light. The analytical  QFI expressions, which are calculated for both SU(2) and SU(1,1) interferometer configuration, helps us understand the dependence of the phase response on the interferometer architecture. Our analysis also examines the role of the lattice structure of non-Gaussian GKP state and demonstrates that the resulting enhancement in phase sensitivity is independent of the specific interferometer architecture.

For the one-parameter (purely damped) finite-energy GKP state, as described in
\eqref{eq:gkp_one_param}, we find that the relative QFI of the GKP as well as squeezed-vacuum enhanced interferometer is governed by the position-quadrature variance of the GKP state. The position quadrature variance depends on the parameter $\Delta$ which determines the peak width and the envelope width. At $\Delta\simeq0.3282$ for the finite-energy GKP state, the QFI of both the states becomes equal. We notice that this crossover is independent of the squeezing strength $\kappa$ and optical transmissivity $\eta$. We also notice that optical loss reduces the relative QFI advantage by introducing vacuum fluctuations into the detected quadrature.

We then perform a similar analysis using the generalized description of the finite energy GKP state by allowing the lattice peak width ($\sigma_{\mathrm{peak}}$) and the Gaussian envelope width ($\sigma_{\mathrm{env}}$) to vary independently (described in \eqref{eq:gkp_two_width}). We observe that the envelope width has a greater effect on the relative QFI. We then determine the relative QFI at fixed mean photon number and observe that the Heisenberg uncertainty principle places an upper bound on the position quadrature variance. From this we can then determine that the QFI of the finite-energy GKP state cannot exceed that of a squeezed-vacuum state with the same mean photon number. 

Thus, our results establish a unified framework for studying finite-energy GKP states in both passive and active interferometric architectures and examine the roles played by squeezing, optical loss, GKP lattice structure, and resource constraints, providing a basis for future investigations of GKP-enhanced interferometry.

A realistic interferometer is inevitably subject to various noise sources and imperfections \cite{Rana_PhysRevD}, including electronic noise, mode mismatch \cite{Squeezing_mode_mismatch}, imperfect state preparation, laser frequency and intensity \cite{Frequency_intensity_noise} and beam-pointing fluctuations \cite{Mueller_05}, environmental disturbances \cite{Nguyen_2021}, and detector inefficiency\cite{progress_in_optics_2015, SCHNABEL20171, LIGO_2016}. Extending the present analysis to incorporate these effects and investigating their impact on the quantum Fisher information (QFI) and phase sensitivity for GKP-state, in comparison with the established results for squeezed-vacuum states \cite{Rana_PhysRevD}, would provide a more comprehensive assessment of the practical metrological advantage and robustness of GKP-states under realistic experimental conditions.

It would also be interesting to investigate whether the error-correction properties of GKP states can be incorporated directly into the sensing protocol. Recent proposals for realizing GKP states on a torus \cite{Sijo_2025}, and related approaches to finite-energy bosonic codes \cite{Bosonic_qauntum_comp,Boson_sampling_2,Sivak2023}, may provide promising routes toward experimentally accessible GKP resources for precision sensing.

\bibliographystyle{apsrev4-2-5authors}
\bibliography{apssamp}


\appendix

\section{From the Ideal GKP State to the Finite-Energy GKP State}
\label{app:qunaught_damping}

The ideal qunaught state consists of an infinite comb of position
eigenstates,
\begin{equation}
\braket{q}{\mathrm{GKP}_{\mathrm{ideal}}}
\propto
\sum_{s=-\infty}^{\infty}
\delta\left(q-s\sqrt{2\pi}\right).
\end{equation}
The delta-function peaks correspond to infinitely well-localized
lattice points and consequently require an unbounded range of photon
numbers. To obtain a physical finite-energy state, the high-energy
components of the ideal state can be suppressed by applying the
non-unitary operator $e^{-\Delta^2\hat{n}}$ where $\hat{n}=\hat{a}^{\dagger}\hat{a}$ and 
$\Delta>0$ controls the strength of the damping. The resulting
state is
\begin{equation}
\ket{\mathrm{GKP}_{\Delta}}
=
\mathcal{N}_{\Delta}
e^{-\Delta^2\hat{n}}
\ket{\mathrm{GKP}_{\mathrm{ideal}}},
\end{equation}
where $\mathcal{N}_{\Delta}$ ensures normalization. To see how this damping modifies the ideal comb, consider its action
in the position basis. Inserting a resolution of the identity gives
\begin{align}
\braket{q}{\mathrm{GKP}_{\Delta}}
&\propto
\sum_{s=-\infty}^{\infty}
\bra{q}e^{-\Delta^2\hat{n}}
\ket{s\sqrt{2\pi}}
\nonumber\\
&=
\sum_{s=-\infty}^{\infty}
K_{\Delta}
\left(q,s\sqrt{2\pi}\right),
\end{align}
where $K_{\Delta}(q,q') = \bra{q}e^{-\Delta^2\hat{n}}\ket{q'}$ is the position-space kernel of the damping operator. Using $\hat{n}  = \frac{\hat{q}^{\,2}+\hat{p}^{\,2}-1}{2}$, 
the operator $e^{-\Delta^2\hat{n}}$ is related to the imaginary-time
propagator of the harmonic oscillator. Its exact position-space kernel is 
\begin{align}
&K_{\Delta}(q,q') \nonumber\\
= &\frac{e^{\Delta^2/2}}
{\sqrt{2\pi\sinh(\Delta^2)}}
\exp\left[ -\frac{ (q^2+q'^2)\cosh(\Delta^2)-2qq'
}{ 2\sinh(\Delta^2) } \right].
\end{align}
Thus, the action of the damping operator replaces each ideal smooth Gaussian-like function while simultaneously suppressing contributions from lattice points far from the origin. In the regime of small $\Delta$, the kernel becomes sharply peaked around $q=q'$. Using $\sinh(\Delta^2)\simeq\Delta^2$ and $\cosh(\Delta^2)\simeq 1$, the dominant dependence of the kernel on the separation between the two coordinates takes the Gaussian form
\begin{align}
&K_{\Delta}(q,q') \nonumber \\
\sim &\exp\left[ -\frac{(q-q')^2}{2\Delta^2}\right]
\times \text{slowly varying envelope factors}.
\end{align}
Consequently, each delta-function peak of the ideal qunaught state is broadened into a Gaussian of characteristic width $\Delta$,
while the damping of the higher-energy lattice components produces a global Gaussian envelope. The resulting finite-energy state is therefore conveniently represented, in the regime relevant to the calculations below, by
the Gaussian-comb form
\begin{align}
&\braket{q}{\mathrm{GKP}_{\Delta}} \nonumber\\
= &
\mathcal{N}_{\Delta}
\exp\left(-\pi\Delta^2q^2\right)
\sum_{s=-\infty}^{\infty}
\exp\left[
-\frac{
\left(q-s\sqrt{2\pi}\right)^2
}{
2\Delta^2
}
\right].
\end{align}
The first factor describes the global suppression of lattice points
away from the origin, whereas the summation describes the individual
Gaussian peaks centred at the ideal lattice positions
$q=s\sqrt{2\pi}$. We express the envelope in the standard Gaussian form,
\begin{equation}
\exp\left(-\pi\Delta^2q^2\right)
= \exp\left(-\frac{q^2} {2\sigma_{\mathrm{env}}^2}\right),
\end{equation}
which gives $\sigma_{\mathrm{env}} = \frac{1}{\sqrt{2\pi}\Delta}$.The width parameter of each individual peak is $\sigma_{\mathrm{peak}}=\Delta$. Thus, within this one-parameter representation, the two characteristic widths satisfy $\sigma_{\mathrm{peak}}\sigma_{\mathrm{env}}
= \frac{1}{\sqrt{2\pi}}$. The Gaussian-comb representation can therefore be understood as the finite-energy approximation to the ideally infinite comb. In the limit $\Delta\rightarrow0$, the individual peaks become increasingly narrow while the envelope becomes increasingly broad, that is, $\sigma_{\mathrm{peak}}\rightarrow0$ and $\sigma_{\mathrm{env}}\rightarrow\infty$ and the state approaches the ideal qunaught comb.

\section{SU(2) Interferometer Transformation} \label{sec:app_SU2}

This appendix derives the input-output relations of the SU(2) interferometer used throughout the paper and establishes the linear relation between a small differential phase shift and the resulting displacement of the dark-port quadrature.

\subsection{Beam Splitter Transformation}

A lossless beam splitter is a passive linear optical element that mixes two bosonic modes while preserving the total photon number. Let $a$ and $b$ denote the input annihilation operators satisfying

\begin{align}
[a,a^\dagger]=[b,b^\dagger]=1, \qquad
[a,b]=[a,b^\dagger]=0. \label{eq:input_operators}
\end{align}

The corresponding output operators are

\begin{align}
a_{\rm out}
    &=ta+rb,\\
b_{\rm out}
    &=-r^*a+t^*b,
\end{align}

where the complex transmission and reflection coefficients satisfy

\begin{align}
|t|^2+|r|^2=1.
\end{align}

The transformation is generated by the unitary operator

\begin{align}
U_{\rm BS} = \exp\!\left[\theta
\left(e^{i\varphi}a^\dagger b - e^{-i\varphi}ab^\dagger
\right) \right],
\end{align}

with

\begin{align}
t=\cos\theta, \qquad r=e^{i\varphi}\sin\theta.
\end{align}

The generators

\begin{align}
J_+=a^\dagger b,\qquad 
J_-=ab^\dagger,\qquad
J_z=\frac12(a^\dagger a-b^\dagger b)
\end{align}

satisfy the $\mathfrak{su}(2)$ algebra, identifying the beam splitter as an $\mathrm{SU}(2)$ rotation in the two-mode Hilbert space.

\subsection{Quadrature convention}

Throughout the appendix we use the optical quadratures

\begin{align}
 X_1 = \frac{ a+ a^\dagger}{2},
\qquad X_2 = \frac{ a- a^\dagger}{2i},
\end{align}

which satisfy

\begin{align}
[ X_1, X_2] = \frac{i}{2},
\end{align}

with vacuum variances

\begin{align}
(\Delta X_1)^2_{\rm vac} = (\Delta X_2)^2_{\rm vac}
= \frac14.
\end{align}

The corresponding canonical quadratures are

\begin{align}
q=\sqrt2\,{X}_1, \qquad p=\sqrt2\,{X}_2,
\end{align}

for which

\begin{align}
[q,p]=i.
\end{align}

Since finite-energy GKP states are defined in the canonical quadratures, the lattice spacing
$\sqrt{\pi}$ corresponds to

\begin{align}
\ell=\sqrt{\frac{\pi}{2}}
\end{align}

in the ${X}_2$ convention adopted throughout this work.

\subsection{Dark-port transformation}

We consider a Mach-Zehnder (Michelson-equivalent) interferometer consisting of two balanced beam splitters separated by a differential phase shift $\phi$. The input state is

\begin{align}
|\Psi_{\rm in}\rangle = |\alpha\rangle_a \otimes |\psi\rangle_b,
\end{align}

where the coherent field enters the bright port and the GKP or squeezed vaccum state is injected into the dark port.

The first beam splitter performs the transformation

\begin{align}
\begin{pmatrix}
a_1\\
b_1
\end{pmatrix}
=
\frac1{\sqrt2}
\begin{pmatrix}
1&i\\
i&1
\end{pmatrix}
\begin{pmatrix}
a\\
b
\end{pmatrix},
\end{align}

followed by a phase shift applied to the first arm transforming it as

\begin{align} 
a_1 \mapsto e^{i\phi}a_1,
\end{align}

and recombination at the second balanced beam splitter,

\begin{align}
\begin{pmatrix}
a_{\rm out}\\
b_{\rm out}
\end{pmatrix}
= \frac1{\sqrt2} 
\begin{pmatrix}
1&-i\\ 
-i&1
\end{pmatrix}
\begin{pmatrix}
e^{i\phi}a_1\\
b_1
\end{pmatrix}.
\end{align}

Substituting the intermediate fields gives

\begin{align}
b_{\rm out} = \frac12 \left[ (e^{i\phi}-1)a
+ (e^{i\phi}+1)b \right].
\label{eq:darkport_exact}
\end{align}

\subsection{Linearized dark-fringe regime}

The interferometer is operated close to the dark fringe,

\begin{align}
|\phi|\ll1,
\end{align}

for which

\begin{align}
e^{i\phi} = 1+i\phi+\mathcal O(\phi^2).
\end{align}

Substituting this expansion into Equation ~\eqref{eq:darkport_exact} yields

\begin{align}
b_{\rm out} = b + \frac{i\phi}{2}(a+b)
+ \mathcal O(\phi^2).
\end{align}

The term proportional to $b$ represents only an infinitesimal phase rotation of the dark-port mode and does not contribute to the leading-order phase signal. Neglecting this local rotation gives

\begin{align}
b_{\rm out} \simeq b + \frac{i\phi}{2}a. \label{eq:darkport_linear}
\end{align}

Equation~\eqref{eq:darkport_linear} shows that a small differential phase shift couples the bright-port field into the dark port.

\subsection{Output quadrature displacement}

For a coherent input field, it is convenient to decompose the annihilation
operator into its mean amplitude and quantum fluctuations,

\begin{align}
a=\alpha+\delta a,
\end{align}

where $\alpha=\langle a\rangle\in\mathbb R$ and
$\langle\delta a\rangle=0$.
Substituting into Equation ~(\ref{eq:darkport_linear}) gives

\begin{align}
b_{\rm out} = b + i\frac{\alpha\phi}{2} + \frac{i\phi}{2}\delta a.
\end{align}

The second term is a classical displacement of the dark-port mode, while the
last term describes the quantum fluctuations of the bright-port field coupled
into the output. In the usual operating regime Michelson
interferometers, the coherent amplitude satisfies $\alpha\gg \delta a$, so the
classical displacement dominates and the fluctuation term provides only a small
quantum correction. Thus the output mode can be approximated as

\begin{align}
b_{\rm out} \simeq b + i\frac{\alpha\phi}{2}.
\end{align}

The corresponding phase quadrature is therefore

\begin{align}
X_{2,\rm out} = X_{2,b} + \frac{\phi}{2}X_{1,a},
\end{align}

which immediately yields

\begin{align}
\langle X_{2,\rm out}\rangle = \frac{\alpha\phi}{2},
\label{eq:su2_response}
\end{align}

since $\langle X_{1,a}\rangle=\alpha$ for a coherent state.

Thus, in the linearized regime, the SU(2) interferometer converts a differential phase shift into a displacement of the dark-port phase quadrature proportional to the coherent carrier amplitude.

\section{SU(1,1) interferometer calculations}\label{sec:app_SU11}

This appendix derives the input-output relations of the SU(1,1) interferometer used in the main text and establishes its linear response to a small differential phase shift. Unlike the passive SU(2) interferometer, the SU(1,1) configuration employs optical parametric amplifiers (OPAs), which amplify the phase signal before readout.

\subsection{Two-mode squeezing transformation}

An SU(1,1) interferometer replaces the beam splitters of a conventional Mach-Zehnder interferometer with two OPAs. The transformation does not conserve total photon number, instead it amplifies quantum fluctuations through stimulated pair creation. The corresponding Bogoliubov transformation between the input modes $a$ and $b$ is

\begin{align}
a_{\rm out} &= ua+vb^\dagger, \\
b_{\rm out} &= ub+va^\dagger,
\end{align}

where

\begin{align}
u=\cosh r, \qquad v=e^{i\varphi}\sinh r,
\end{align}

with $r$ denoting the parametric gain. Preservation of the canonical commutation relations shown in \eqref{eq:input_operators}, requires

\begin{align}
|u|^2-|v|^2=1.
\end{align}

The transformation is generated by the two-mode squeezing operator

\begin{align}
U_{\rm OPA} = \exp\!\left[ r \left( e^{i\varphi}a^\dagger b^\dagger - e^{-i\varphi}ab \right)\right],
\end{align}

whose generators

\begin{align} 
K_+=a^\dagger b^\dagger, \qquad K_-=ab, \qquad
K_0=\frac12(a^\dagger a+b^\dagger b+1)
\end{align}

satisfy the $\mathfrak{su}(1,1)$ algebra.

\subsection{Dark-port transformation}

The interferometer consists of two identical OPAs separated by a differential phase shift $\phi$ applied to mode $a$. The first OPA performs the transformation

\begin{align}
a_1 &= \cosh r\,a + \sinh r\,b^\dagger, \\
b_1 &= \cosh r\,b + \sinh r\,a^\dagger.
\end{align}

After the phase shift,

\begin{align}
a_2=e^{i\phi}a_1, \qquad b_2=b_1,
\end{align}

the second OPA is operated with the opposite pump phase so that it exactly inverts the first transformation when $\phi=0$. The output dark-port operator is therefore

\begin{align}
b_{\rm out} = \cosh r\,b_2 - \sinh r\,a_2^\dagger,
\end{align}

which simplifies to

\begin{align}
b_{\rm out} = u(\phi)b + v(\phi)a^\dagger, \label{eq:su11_exact}
\end{align}

with

\begin{align}
u(\phi) &= \cosh^2r - e^{-i\phi}\sinh^2r, \\
v(\phi) &= \sinh r\cosh r \left( 1-e^{-i\phi} \right).
\end{align}

At the operating point ($\phi = 0$),

\begin{align} 
u(0)=1, \qquad v(0)=0,
\end{align}

so that

\begin{align}
b_{\rm out}=b.
\end{align}

Thus, the second OPA exactly de-amplifies the interferometer at the dark fringe.

\subsection{Linearized dark-fringe regime}

For small phase excursions,

\begin{align}
|\phi|\ll1,
\end{align}

the phase factor may be expanded as

\begin{align}
e^{-i\phi} = 1-i\phi+\mathcal O(\phi^2).
\end{align}

Using this expansion,

\begin{align}
u(\phi) &= 1+i\phi\sinh^2r + \mathcal O(\phi^2), \\
v(\phi) &= i\phi\sinh r\cosh r + \mathcal O(\phi^2).
\end{align}

Introducing the nonlinear gain

\begin{align}
G = \sinh r\cosh r = \frac12\sinh(2r),
\end{align}

the output field becomes

\begin{align}
b_{\rm out} \simeq \left( 1+i\phi\sinh^2r
\right)b + iG\phi\,a^\dagger. \label{eq:su11_linear}
\end{align}

The first term corresponds to an infinitesimal phase rotation of the dark port input field, while the second contains the measurable phase signal amplified by the nonlinear gain.

\subsection{Output quadrature response}

Assume the bright input port contains a coherent state of real amplitude $\alpha$,

\begin{align}
|\alpha\rangle_a,
\end{align}

while the state injected into the dark port satisfies

\begin{align}
\langle b\rangle=0.
\end{align}

Taking the expectation value of Equation ~\eqref{eq:su11_linear} gives

\begin{align}
\langle b_{\rm out}\rangle
= iG\alpha\phi.
\end{align}

To leading order, the phase signal therefore appears as a displacement along the phase quadrature,

\begin{align}
\langle X_{2,\rm out}\rangle
= G\alpha\phi,
\label{eq:su11_response}
\end{align}

with local phase responsivity

\begin{align}
\frac{\partial}{\partial\phi}
\langle X_{2,\rm out}\rangle
= G\alpha.
\end{align}

Since the interferometer is operated close to the dark fringe, the unknown phase is treated as a small perturbation about the operating point. Corrections to the covariance matrix therefore enter only at $\mathcal O(\delta\phi^2)$. Consequently, the output quadrature variance is independent of the phase to leading order,

\begin{align}
\mathrm{Var}(X_{2,\rm out}) = \mathrm{Var}(X_{2}) + \mathcal O(\phi^2),
\end{align}

which justifies treating the variance as constant in the local Fisher-information analysis in later Appendices.

\subsection{High-gain limit}

For large parametric gain,

\begin{align}
r\gg1,
\end{align}

the nonlinear amplification factor satisfies

\begin{align}
G = \sinh r\cosh r = \frac14e^{2r} \left(
1+\mathcal O(e^{-4r})\right),
\end{align}

demonstrating the exponential enhancement of the local phase responsivity with increasing OPA gain.

\section{Quantum Fisher Information Without Optical Loss}\label{sec:app_QFI_noloss}

This appendix derives the quantum Fisher information (QFI) for both momentum-squeezed finite-energy GKP states and momentum-squeezed vacuum states. Since the SU(2) and SU(1,1) interferometers differ only in the proportionality constant relating the unknown phase shift to the output quadrature displacement, both architectures admit a common QFI derivation.

\subsection{Phase encoding}

From the linearized input-output relations derived in Appendices~A and B, the dark-port field may be written as

\begin{align}
a_{\rm out}
\simeq
a_{\rm in}
+
iA\phi,
\end{align}

where

\begin{align}
A=
\begin{cases}
\dfrac{\alpha}{2},
&
\mathrm{SU(2)},
\\[2mm]
G\alpha,
&
\mathrm{SU(1,1)},
\end{cases}\label{eq:cases_A}
\end{align}

where $\alpha$ is the coherent carrier amplitude and
$G=\sinh r \cosh r$ denotes the nonlinear amplification factor of the SU(1,1) interferometer. Thus, to first order in the phase shift,

\begin{align}
\delta\alpha = iA\phi,
\end{align}

corresponding to a displacement along the optical phase quadrature.

\subsection{Effective Generator}

Using the canonical quadrature decomposition

\begin{align}
a=\frac{q+ip}{\sqrt2},
\end{align}

the phase-induced displacement of the coherent amplitude

\begin{align}
\delta\alpha=iA\phi
\end{align}

corresponds to the quadrature displacements

\begin{align}
\delta q = 0, \qquad \delta p = \sqrt2A\phi.
\end{align}

Translations along the momentum direction are generated by the conjugate position quadrature. The phase encoding is therefore described by the displacement operator

\begin{align}
\hat D(\phi) = \exp\!\left( -i\sqrt2A\phi\,q \right),
\end{align}

from which the effective generator immediately follows,

\begin{align}
H_{\rm eff} = \sqrt2Aq. \label{eq:Heff}
\end{align}

\subsection{Generic Quantum Fisher Information}

The dark-port state is either the squeezed vacuum or finite-energy GKP state. In the absence of optical loss, this is a pure state undergoing the unitary evolution
\begin{align}
|\psi(\phi)\rangle = e^{-iH_{\rm eff}\phi}|\psi\rangle.
\end{align}
For a pure dark port input state, the quantum Fisher information reduces to
\begin{align}
F_Q = 4\,\mathrm{Var}(H_{\rm eff}),
\end{align}
where the variance is evaluated with respect to the input state.
Using Equation ~(\ref{eq:Heff}), this becomes

\begin{align}
F_Q = 8A^2 \mathrm{Var}(q).
\label{eq:generic_qfi}
\end{align}

Equation~(\ref{eq:generic_qfi}) is independent of the specific interferometer geometry and shows that the QFI depends only on the variance
of the generator $H_{\rm eff}\propto q$. Since the unknown phase is encoded as a displacement of the momentum quadrature, increasing the variance of the conjugate position quadrature enhances the achievable precision. Consequently, momentum-squeezed states are natural candidates for phase estimation. The following sections evaluate momentum-squeezed vacuum states and momentum-squeezed finite-energy GKP states.

\subsection{Finite-Energy GKP States}
\label{eq:qfi_gkp_noloss}

We now derive the quantum Fisher information for finite-energy GKP
states. 

\subsubsection{Variance of the Position Quadrature as function of $\Delta$}
\label{app:gkp_variance}

We evaluate the variance of the position quadrature for an unsqueezed finite-energy GKP state. This provides the starting
point for the subsequent analysis of the squeezed state. For $\kappa=0$, the finite-energy qunaught wavefunction can be written as
\begin{align}
\psi(q) &= \mathcal{N}_{\Delta} \exp\left(
-\pi\Delta^2 q^2 \right) \sum_{s=-\infty}^{\infty}
\exp\left[ -\frac{ \left(q-s\sqrt{2\pi}\right)^2
}{ 2\Delta^2 } \right]. \label{eq:unsqueezed_qunaught_wavefunction}
\end{align}

Taking the modulus squared gives
\begin{align}
|\psi(q)|^2 =&\mathcal{N}_{\Delta}^2 \exp\left( -2\pi\Delta^2 q^2 \right) \cdot \nonumber \\
& \sum_{s,t=-\infty}^{\infty} \exp\left[ -\frac{ \left(q-s\sqrt{2\pi}\right)^2 + \left(q- t\sqrt{2\pi}\right)^2 }{ 2\Delta^2 } \right].
\label{eq:exact_unsqueezed_gkp_probability}
\end{align}

The terms with $s=t$ describe the contributions  from individual lattice components, while the terms with $s\neq t$ describe their overlap and interference. The state is symmetric under the transformation $q\mapsto -q$, and hence $\langle q\rangle=0$. The position variance is therefore $\mathrm{Var}(q)
= \langle q^2\rangle$. For a given pair of lattice indices $(s,t)$, the exponent appearing
in Equation ~(\ref{eq:exact_unsqueezed_gkp_probability}) is
\begin{align}
-2\pi\Delta^2q^2
-\frac{
(q-s\sqrt{2\pi})^2
+
(q-t\sqrt{2\pi})^2
}{
2\Delta^2
}.
\end{align}

We introduce the quantity $D=1+2\pi\Delta^4$. The exponent can then be written as
\begin{align}
&-\frac{D}{\Delta^2}
\left(
q-\mu_{st}
\right)^2
-\frac{\pi(s-t)^2}{2\Delta^2}
-\frac{\pi^2\Delta^2(s+t)^2}{D},
\label{eq:unsqueezed_gkp_completed_square}
\end{align}
where
\begin{equation}
\mu_{st}
=
\frac{\sqrt{2\pi}(s+t)}{2D}.
\label{eq:unsqueezed_gkp_pair_mean}
\end{equation}

Thus, for each pair $(s,t)$, the integral over $q$ is Gaussian, with mean $\mu_{st}$ and variance $\sigma_{st}^2 = \frac{\Delta^2}{2D}$. The terms in the exponent independent of $q$ become the corresponding weight of the pair $(s,t)$, that is:
\begin{equation}
W_{st}
=
\exp\left[
-\frac{\pi(s-t)^2}{2\Delta^2}
-\frac{\pi^2\Delta^2(s+t)^2}{D}
\right].
\label{eq:unsqueezed_gkp_pair_weight}
\end{equation}

The normalization integral is consequently proportional to
\begin{align}
\mathcal{Z}_{\Delta}
=
\sum_{s,t=-\infty}^{\infty}
W_{st}.
\label{eq:unsqueezed_gkp_normalization_sum}
\end{align}
The common factors arising from the Gaussian integral, together
with the normalization constant $\mathcal{N}_{\Delta}$, cancel from
normalized expectation values. The second moment can be evaluated using the standard Gaussian
identity
\begin{equation}
\int_{-\infty}^{\infty}
dq\,q^2
e^{-A(q-\mu)^2}
=
\sqrt{\frac{\pi}{A}}
\left(
\mu^2+\frac{1}{2A}
\right).
\end{equation}

Using the expressions of $\mu_{st}$ and $W_{st}$, we obtain
\begin{align}
\langle q^2\rangle
&=
\frac{\Delta^2}{2D}
+
\frac{
\displaystyle
\sum_{s,t=-\infty}^{\infty}
\mu_{st}^2 W_{st}
}{
\displaystyle
\sum_{s,t=-\infty}^{\infty}
W_{st}
}=
\frac{\Delta^2}{2D}
+
\frac{\pi}{2D^2} \times \nonumber\\
&\frac{
\displaystyle
\sum_{s,t=-\infty}^{\infty}
(s+t)^2
\exp\left[
-\frac{\pi(s-t)^2}{2\Delta^2}
-\frac{\pi^2\Delta^2(s+t)^2}{D}
\right]
}{
\displaystyle
\sum_{s,t=-\infty}^{\infty}
\exp\left[
-\frac{\pi(s-t)^2}{2\Delta^2}
-\frac{\pi^2\Delta^2(s+t)^2}{D}
\right]
}.
\label{eq:unsqueezed_gkp_exact_q2}
\end{align}

Since $\langle q\rangle=0$, the position variance becomes,
\begin{align}
&\mathrm{Var}(q) = \frac{\Delta^2}{2D} + \frac{\pi}{2D^2} \times \nonumber \\
&\frac{ \displaystyle \sum_{s,t=- \infty}^{\infty} (s+t)^2 \exp\left[ -\frac{\pi(s-t)^2}{2\Delta^2} -\frac{\pi^2\Delta^2(s+t)^2}{D}
\right] }{ \displaystyle \sum_{s,t=- \infty}^{\infty} \exp\left[ -\frac{\pi(s-t)^2}{2\Delta^2} -\frac{\pi^2\Delta^2(s+t)^2}{D}
\right] }, \nonumber \\
& \textrm{where} \quad D=1+2\pi\Delta^4.
\label{eq:unsqueezed_gkp_exact_variance}
\end{align}

Equation~(\ref{eq:unsqueezed_gkp_exact_variance}) is the exact position variance of the unsqueezed finite-energy qunaught within the finite-energy wavefunction defined in Equation ~(\ref{eq:unsqueezed_qunaught_wavefunction}). Squeezing can subsequently be reintroduced independently through the transformation of the
position quadrature, $q \mapsto e^{\kappa}q$ which gives $\mathrm{Var}_{\kappa}(q) = e^{2\kappa} \mathrm{Var} (q)$. 

The exact position variance expression in
Equation ~(\ref{eq:unsqueezed_gkp_exact_variance}) can be reduced systematically by imposing the narrow-peak and many-lattice-peaks conditions in succession. This also makes explicit the  approximations underlying the commonly used asymptotic expression. 

In the narrow-peak regime, the width of the individual peaks should be much smaller than the lattice distance, that is, $\Delta \ll \sqrt{2\pi}$. In Equation ~(\ref{eq:unsqueezed_gkp_exact_variance}), the dependence on the difference between the two lattice indices is contained in the factor $\exp\left[ -\frac{\pi(s-t)^2}{2\Delta^2}
\right]$. When $\Delta$ is sufficiently small, all terms with $s\neq t$ are therefore exponentially suppressed. The dominant contribution comes
from the diagonal terms $s=t$. For these terms, $s-t=0$ and $s+t=2s$. Consequently, Equation ~(\ref{eq:unsqueezed_gkp_exact_variance}) reduces to
\begin{align}
\mathrm{Var}(q)
&\simeq
\frac{\Delta^2}{2D}
+
\frac{\pi}{2D^2}
\frac{
\displaystyle
\sum_{s=-\infty}^{\infty}
(2s)^2
\exp\left(
-\frac{4\pi^2\Delta^2s^2}{D}
\right)
}{
\displaystyle
\sum_{s=-\infty}^{\infty}
\exp\left(
-\frac{4\pi^2\Delta^2s^2}{D}
\right)
}
\nonumber\\
&=
\frac{\Delta^2}{2\left(1+2\pi\Delta^4\right)}
+
\frac{2\pi}{\left(1+2\pi\Delta^4\right)^2} \times \nonumber \\
&\frac{
\displaystyle
\sum_{s=-\infty}^{\infty}
s^2
\exp\left(
-\frac{4\pi^2\Delta^2s^2}{1+2\pi\Delta^4}
\right)
}{
\displaystyle
\sum_{s=-\infty}^{\infty}
\exp\left(
-\frac{4\pi^2\Delta^2s^2}{1+2\pi\Delta^4}
\right)
},
\;
D=1+2\pi\Delta^4.
\label{eq:narrow_peak_variance}
\end{align}

For the many-lattice-peaks condition, the width of the envelope should be much larger than the lattice distance which gives $\Delta \ll \frac{1}{2 \pi}$.  The Gaussian weight in the lattice index $s$ appearing in
Equation ~(\ref{eq:narrow_peak_variance}) has an effective width $\sigma_s \sim \frac{\sqrt{D}}{2\pi\Delta}$. Thus, many lattice sites contribute when $\frac{\sqrt{D}}{2\pi\Delta} \gg 1$. In the many-lattice-peaks regime, $\Delta \ll 1$, so that $D = 1 + 2\pi\Delta^4 \simeq 1$. Equation~(\ref{eq:narrow_peak_variance}) therefore becomes
\begin{align}
\mathrm{Var}(q)
&\simeq
\frac{\Delta^2}{2}
+
2\pi
\frac{
\displaystyle
\sum_{s=-\infty}^{\infty}
s^2
e^{-4\pi^2\Delta^2s^2}
}{
\displaystyle
\sum_{s=-\infty}^{\infty}
e^{-4\pi^2\Delta^2s^2}
}.
\label{eq:many_lattice_discrete_variance}
\end{align}

Because many lattice sites contribute, the discrete sums can now be
approximated by Gaussian integrals:
\begin{align}
\sum_{s=-\infty}^{\infty}
e^{-4\pi^2\Delta^2s^2}
&\simeq
\int_{-\infty}^{\infty}
e^{-4\pi^2\Delta^2s^2}\,ds,
\\
\sum_{s=-\infty}^{\infty}
s^2e^{-4\pi^2\Delta^2s^2}
&\simeq
\int_{-\infty}^{\infty}
s^2e^{-4\pi^2\Delta^2s^2}\,ds.
\end{align}
Using the standard Gaussian integrals
\begin{align}
\int_{-\infty}^{\infty}
e^{-as^2}\,ds
&=
\sqrt{\frac{\pi}{a}},
\\
\int_{-\infty}^{\infty}
s^2e^{-as^2}\,ds
&=
\frac{\sqrt{\pi}}{2a^{3/2}},
\end{align}
with
\begin{equation}
a=4\pi^2\Delta^2,
\end{equation}
we obtain
\begin{align}
\frac{
\displaystyle
\int_{-\infty}^{\infty}
s^2e^{-4\pi^2\Delta^2s^2}\,ds
}{
\displaystyle
\int_{-\infty}^{\infty}
e^{-4\pi^2\Delta^2s^2}\,ds
}
&=
\frac{1}{8\pi^2\Delta^2}.
\end{align}
Substitution into Equation ~(\ref{eq:many_lattice_discrete_variance})
therefore gives
\begin{align}
\mathrm{Var}(q)
&\simeq
\frac{\Delta^2}{2}
+
2\pi
\left(
\frac{1}{8\pi^2\Delta^2}
\right)
\nonumber\\
&=
\frac{\Delta^2}{2}
+
\frac{1}{4\pi\Delta^2}
\nonumber\\
&=
\frac{1}{2}
\left(
\Delta^2+
\frac{1}{2\pi\Delta^2}
\right).
\label{eq:unsqueezed_gkp_asymptotic_variance}
\end{align}

Thus, Eq \eqref{eq:unsqueezed_gkp_asymptotic_variance}  follows from the exact position variance by using the narrow-peak approximation which suppresses the off-diagonal terms $s\neq t$ and the many-lattice-peaks approximation which allows the remaining discrete lattice sums to be replaced by Gaussian integrals.

\subsubsection{Numerically approximating the infinite sum} \label{app:numerical_approx}

The exact variance in Equation~(\ref{eq:unsqueezed_gkp_exact_variance}) contains
a double sum $W_{st}$ over the two lattice indices $s$ and $t$.
\begin{align}
W_{st}
=
&\exp\left[
-\frac{\pi(s-t)^2}{2\Delta^2}
-\frac{\pi^2\Delta^2(s+t)^2}{\left(1+2\pi\Delta^4\right)}
\right], 
\label{eq:results_pair_weight}
\end{align}
We introduce the variables $u=s+t$ and $v=s-t$. In terms of these variables, the pair weight factorizes as
\begin{equation}
W_{uv}
=
\exp\left(
-\frac{\pi v^2}{2\Delta^2}
\right)
\exp\left(
-\frac{\pi^2\Delta^2u^2}{1+2\pi\Delta^4}
\right).
\label{eq:factorized_pair_weight}
\end{equation}
Thus, the two factors describe the extent of the pair-weight
distribution along the two lattice directions $v=s-t$ and
$u=s+t$, respectively. These factors  determine how rapidly the terms in the exact double sum
decay as the lattice indices are varied. To identify the corresponding characteristic widths, we compare each
factor with the standard Gaussian form $\exp\left(-{x^2}/{2\sigma^2}\right)$. For the coordinate $v=s-t$, we have $\exp\left(-{\pi v^2}/{2\Delta^2}
\right) = \exp\left(-{v^2}/ {2\sigma_v^2}\right)$ which gives $
\sigma_v={\Delta}/{\sqrt{\pi}}$. 
This factor determines how strongly terms with $s\neq t$ are suppressed. In particular, increasing $|s-t|$ moves away from the diagonal $s=t$, and the corresponding contribution is suppressed on
the scale $\sigma_v$.

For the coordinate $u=s+t$, the second factor can similarly be written as $\exp\left( -{\pi^2\Delta^2u^2}/{\left(1+2\pi\Delta^4\right)}
\right) = \exp\left( -{u^2}/{2\sigma_u^2} \right)$ which gives $\sigma_u = {\sqrt{\left(1+2\pi\Delta^4\right)}}/{\sqrt{2}\pi\Delta}$. This factor determines the range of values of $s+t$ that contribute appreciably to the double sum. In contrast to $\sigma_v$, which controls the suppression of off-diagonal pairs, $\sigma_u$ controls
the extent of the contributing lattice region along the direction
$s+t$.

Thus the width $\sigma_v$ determines how
rapidly the overlap terms decay as the two lattice indices separate, whereas $\sigma_u$ determines the extent of the lattice-index distribution along the direction in which the two indices vary
together. Their dependence on $\Delta$ is also opposite: $\sigma_v\propto\Delta$ and $\sigma_u\propto {\sqrt{\left(1+2\pi\Delta^4\right)}}/{\Delta}$. 
Consequently, depending on the value of $\Delta$, either the $s-t$ or the $s+t$ direction can have the broader Gaussian width.

These widths provide a natural basis for choosing the numerical
summation range. Since the double sum is formally infinite, it was
truncated to $s,t\in[-N,N]$ with the cutoff chosen adaptively as $N=
\left\lceil 8\max(\sigma_u,\sigma_v)
\right\rceil+2$. The use of the larger of $\sigma_u$ and $\sigma_v$ ensures that the
summation range is sufficiently large to capture the broader direction of the pair-weight distribution. The additional $+2$
provides a small numerical safety margin. This procedure is a numerical truncation of the formally infinite sums and does not
correspond to imposing either the narrow-peak or the many-lattice-peaks approximation on the analytical expression.

At a distance of eight standard deviations from the centre, a Gaussian factor of the form $\exp\left(-x^2/(2\sigma^2)\right)$ has the value $\exp\left(-{8^2}/{2}\right) \simeq 1.3\times10^{-14}$. Consequently, the terms associated with the corresponding Gaussian tails are negligibly small at the numerical precision relevant to
the present calculation. The numerical results therefore retain the
full structure of the exact double sum while replacing only the mathematically infinite summation range by a sufficiently large finite one.

\subsubsection{Variance of the Position Quadrature as a function of
Independent Peak and Envelope Widths}
\label{app:gkp_variance_two_parameters}

We now derive the position-quadrature variance for a more general
finite-energy GKP qunaught state in which the width of each lattice
peak and the width of the global envelope are treated as independent
parameters. The position-space wavefunction is taken to be
\begin{align}
\braket{q}{\mathrm{GKP}_{\sigma_{\mathrm{peak}},
\sigma_{\mathrm{env}}}} =
\mathcal{N}_{\sigma_{\mathrm{peak}},\sigma_{\mathrm{env}}}
\exp\left(
-\frac{q^2}{2\sigma_{\mathrm{env}}^2}
\right) \times \nonumber \\
\sum_{s=-\infty}^{\infty}
\exp\left[ 
-\frac{ \left(q-s\sqrt{2\pi}\right)^2 }{
2\sigma_{\mathrm{peak}}^2 }
\right].
\label{eq:gkp_two_parameter_state}
\end{align}
Here, $\sigma_{\mathrm{peak}}$ controls the width of the individual
Gaussian peaks, while $\sigma_{\mathrm{env}}$ controls the extent of
the global Gaussian envelope. Unlike the single-parameter
parameterization, these two widths can therefore be varied
independently.

The state is symmetric under the transformation $q\mapsto -q$.
Thus $\langle q\rangle=0$ and the position variance is simply $\mathrm{Var}(q)=\langle q^2\rangle$. Taking the modulus squared of Eq.~(\ref{eq:gkp_two_parameter_state}) gives
\begin{align}
|\psi(q)|^2 &=
\mathcal{N}_{\sigma_{\mathrm{peak}},\sigma_{\mathrm{env}}}^2
\exp\left( -\frac{q^2}{\sigma_{\mathrm{env}}^2} \right) \nonumber\\
&\quad \times \sum_{s,t=-\infty}^{\infty} 
\exp\left[ -\frac{ \left(q-s\sqrt{2\pi}\right)^2 + \left(q- t\sqrt{2\pi}\right)^2}{2\sigma_{\mathrm{peak}}^2}\right].
\label{eq:gkp_two_parameter_probability}
\end{align}

The diagonal terms with $s=t$ describe the individual lattice
components, while the off-diagonal terms with $s\neq t$ describe the
overlap and interference between different lattice components. For a fixed pair of lattice indices $(s,t)$, the exponent appearing in Eq.~(\ref{eq:gkp_two_parameter_probability})
is then

\begin{equation}
    \begin{aligned}
        &-\frac{q^2}{\sigma_{\mathrm{env}}^2}
        -\frac{(q-\sqrt{2\pi}s)^2+(q-\sqrt{2\pi}t)^2}
        {2\sigma_{\mathrm{peak}}^2}\\
        = & -\frac{q^2}{\sigma_{\mathrm{env}}^2}
        -\frac{2q^2 -2\sqrt{2\pi}q(s+t) + 2\pi(s^2+t^2)}
        {2\sigma_{\mathrm{peak}}^2}\\
        =& -\left(\frac{1}{\sigma_{\mathrm{env}}^2} + \frac{1}{\sigma_{\mathrm{peak}}^2}\right)q^2 + \frac{\sqrt{2\pi}(s+t)}{\sigma_{\mathrm{peak}}^2}q - \frac{2\pi(s^2+t^2)}
        {2\sigma_{\mathrm{peak}}^2}\\
        =&-\frac{\left(\sigma_{\mathrm{peak}}^2 + \sigma_{\mathrm{env}}^2\right)}{\sigma_{\mathrm{peak}}^2
        \sigma_{\mathrm{env}}^2}
        \left(q-\mu_{st}\right)^2 -\frac{\pi(s-t)^2}
        {2\sigma_{\mathrm{peak}}^2} \\
        &-\frac{\pi(s+t)^2}
        {2\left(\sigma_{\mathrm{peak}}^2 + \sigma_{\mathrm{env}}^2\right)}
    \end{aligned}
\end{equation}

where the center of the Gaussian associated with the pair $(s,t)$ is
\begin{equation}
\mu_{st}
=
\frac{\sqrt{2\pi}\,\sigma_{\mathrm{env}}^2
(s+t)}{\left(\sigma_{\mathrm{peak}}^2 + \sigma_{\mathrm{env}}^2\right)}.
\end{equation}

with Gaussian variance
\begin{equation}
\sigma_{st}^2 =  \frac{ \sigma_{\mathrm{peak}}^2
\sigma_{\mathrm{env}}^2
}{
2\left(\sigma_{\mathrm{peak}}^2 + \sigma_{\mathrm{env}}^2\right)
}.
\label{eq:gkp_two_parameter_pair_variance}
\end{equation}

The remaining part of the exponent determines the weight associated
with the pair $(s,t)$. We define
\begin{equation}
W_{st}
=
\exp\left[
-\frac{\pi(s-t)^2}
{2\sigma_{\mathrm{peak}}^2}
-\frac{\pi(s+t)^2}
{2\Sigma^2}
\right].
\label{eq:gkp_two_parameter_pair_weight}
\end{equation}
Thus 
\begin{equation}
    |\psi(q)|^2 =\mathcal{N}_{\sigma_{\mathrm{peak}},\sigma_{\mathrm{env}}}^2\exp \left(\frac{\left(q - \mu_{st}\right)^2}{2\sigma_{st}^2} W_{st}\right)
\end{equation}
The normalization integral can therefore be written as
\begin{align}
\int_{-\infty}^{\infty}dq\,|\psi(q)|^2
=
&\mathcal{N}_{\sigma_{\mathrm{peak}},\sigma_{\mathrm{env}}}^2
\sqrt{\frac{\pi}{2\sigma_{st}^2}} \mathcal{Z}, \nonumber \\
&  \textrm{where} \quad \mathcal{Z} = \sum_{s,t=-\infty}^{\infty}W_{st}
\end{align}

Consequently,
\begin{align}
\langle q^2\rangle  &= \frac{\sigma_{\mathrm{peak}}^2
\sigma_{\mathrm{env}}^2}{2\left(\sigma_{\mathrm{peak}}^2 + \sigma_{\mathrm{env}}^2\right)} \nonumber
\\&+\frac{\pi\sigma_{\mathrm{env}}^4}{2\left(\sigma_{\mathrm{peak}}^2 + \sigma_{\mathrm{env}}^2\right)^2
}\frac{\displaystyle \sum_{s,t=-\infty}^{\infty}(s+t)^2
W_{st}}{\displaystyle \sum_{s,t=-\infty}^{\infty}W_{st}}.
\label{eq:gkp_two_parameter_q2}
\end{align}

Using Eq.~(\ref{eq:gkp_two_parameter_pair_weight}), the position
variance can therefore be written explicitly as
\begin{align}
&\mathrm{Var}(q)
= \frac{\sigma_{\mathrm{peak}}^2\sigma_{\mathrm{env}}^2}{2\left(\sigma_{\mathrm{peak}}^2+\sigma_{\mathrm{env}}^2\right)} + \frac{
\pi\sigma_{\mathrm{env}}^4}{
2\left(\sigma_{\mathrm{peak}}^2 + \sigma_{\mathrm{env}}^2\right)^2} \nonumber\\
&\frac{
\displaystyle
\sum_{s,t=-\infty}^{\infty}
(s+t)^2
\exp\left[
-\frac{\pi(s-t)^2}
{2\sigma_{\mathrm{peak}}^2}
-\frac{\pi(s+t)^2}
{2\left(
\sigma_{\mathrm{peak}}^2
+
\sigma_{\mathrm{env}}^2
\right)}
\right]
}{
\displaystyle
\sum_{s,t=-\infty}^{\infty}
\exp\left[
-\frac{\pi(s-t)^2}
{2\sigma_{\mathrm{peak}}^2}
-\frac{\pi(s+t)^2}
{2\left(
\sigma_{\mathrm{peak}}^2
+
\sigma_{\mathrm{env}}^2
\right)}
\right]
}.
\label{eq:gkp_two_parameter_exact_variance}
\end{align}

Equation~(\ref{eq:gkp_two_parameter_exact_variance}) is the 
position variance for the generalized finite-energy GKP state within the wavefunction model of Eq.~(\ref{eq:gkp_two_parameter_state}). The first term in Eq.~(\ref{eq:gkp_two_parameter_exact_variance})
originates from the intrinsic Gaussian width associated with each $(s,t)$ pair. The second term accounts for the distribution of the effective pair centers $\mu_{st}$ over the lattice and therefore contains the dependence on the global envelope and the lattice structure. The two contributions cannot, in general, be interpreted
as independent peak and lattice-center variances because the off-diagonal terms modify both the pair weights and the effective Gaussian width.

For the special parameterization
\begin{equation}
\sigma_{\mathrm{peak}} = {\Delta}, \qquad \sigma_{\mathrm{env}} = \frac{1}{\sqrt{2\pi}\Delta},
\end{equation}
the generalized state reduces to the previously considered one-parameter finite-energy GKP state. Equation~(\ref{eq:gkp_two_parameter_exact_variance}) therefore provides the natural generalization of the earlier variance expression while allowing the peak width and envelope width to be varied independently.
\subsubsection{Quantum Fisher Information} 

Substituting the position variance, Equation ~\eqref{eq:unsqueezed_gkp_exact_variance} , into the generic QFI expression, Equation ~(\ref{eq:generic_qfi}), we get the quantum Fisher information of squeezed finite-energy GKP states as 
\begin{align}
   F_Q = 8A^2 {e^{2\kappa}} \mathrm{Var}(q)\label{eq:gkp_qfi_lossless}
\end{align}
 where the interferometer-dependent scale factor $A$ is defined in Equation ~(\ref{eq:cases_A}). Hence,

\begin{align}
F_Q^{\rm SU(2)} = 2 \alpha^2 {e^{2\kappa}} \mathrm{Var}(q),
\end{align}

while

\begin{align}
F_Q^{\rm SU(1,1)} = 8 G^2 \alpha^2 {e^{2\kappa}}
    \mathrm{Var}(q).
\end{align}

The nonlinear amplification of the SU(1,1) interferometer therefore enhances the QFI by the factor $4G^2$ relative to the passive SU(2) configuration.

\subsection{Squeezed Vacuum States}

We now derive the quantum Fisher information for squeezed vacuum states. 

\subsubsection{Variance of the Position Quadrature}

For a momentum-squeezed vacuum state with squeezing parameter $\kappa$, the
quadrature variances are

\begin{align}
\operatorname{Var}(q) &= \frac12e^{2\kappa},\\
\operatorname{Var}(p) &= \frac12e^{-2\kappa}.
\end{align}

\subsubsection{Quantum Fisher Information}
Substituting the position variance into the generic QFI expression
gives

\begin{align}
F_Q = 8A^2 \left( \frac12e^{2\kappa} \right)
= 4A^2e^{2\kappa} \label{eq:sq_qfi_lossless}.
\end{align}
where $A$ is as in \eqref{eq:cases_A}. Hence,

\begin{align}
F_Q^{\rm SU(2)} = \alpha^2 {e^{2\kappa}},
\end{align}

and

\begin{align} 
F_Q^{\rm SU(1,1)} = 4G^2 \alpha^2 {e^{2\kappa}}.\label{eq:sq_qfi}
\end{align}

Thus, as in the finite-energy GKP case, the distinction between the SU(2) and SU(1,1) interferometers appears only through the scale factor $A$, with the active SU(1,1) configuration enhancing the QFI by a factor of $4G^2$.

\section{Approximate Quantum Fisher Information With Optical Loss}\label{sec:app_QFI_loss}

This appendix derives approximate expressions for the quantum Fisher information (QFI) in the presence of optical loss. While the exact QFI of the mixed states produced by loss is generally difficult to evaluate analytically, particularly for non-Gaussian states such as finite-energy GKP states, the framework developed in Appendix~C provides a simple approximation by propagating the quadrature variance through a Gaussian loss channel.

\subsection{Gaussian Loss Channel}

Optical loss is modelled as a beam splitter of power transmissivity
$\eta$, which mixes the input mode with an environmental vacuum mode.
The quadrature operators therefore transform as

\begin{align}
q &\mapsto \sqrt{\eta}\,q + \sqrt{1-\eta}\,q_v, \\
p &\mapsto \sqrt{\eta}\,p + \sqrt{1-\eta}\,p_v,
\end{align}

where the vacuum quadratures satisfy

\begin{align}
\operatorname{Var}(q_v) = \operatorname{Var}(p_v) = \frac12.
\end{align}

Since the input and vacuum modes are initially uncorrelated, the
quadrature variances transform according to

\begin{align}
{\operatorname{Var}(q)_\eta = \eta\,\operatorname{Var}(q)
+ \frac{1-\eta}{2},}
\label{eq:loss_variance}
\end{align}

with an identical expression for the momentum quadrature.

\subsection{Finite-Energy GKP States}
We first evaluate the effect of optical loss on the finite-energy GKP state. Using the lossless variance derived in Equation \eqref{eq:unsqueezed_gkp_exact_variance}, the variance after optical loss becomes

\begin{align}
{\operatorname{Var}(q)_\eta
\simeq \eta e^{2\kappa} \mathrm{Var}(q)
+ \frac{1-\eta}{2}.}
\label{eq:gkp_loss_variance1}
\end{align}

Substituting Equation ~(\ref{eq:gkp_loss_variance1}) into the effective-displacement approximation gives,

\begin{align}
F_Q^{(\eta)} \approx 8A^2 \left[\eta e^{2\kappa} \mathrm{Var}(q) + \frac{1-\eta}{2}\right].\label{eq:loss_gkp_qfi}
\end{align}
where $A$ is the interferometer-dependent scale factor defined in \eqref{eq:cases_A}.

\subsection{ Squeezed Vacuum States}

Using the Gaussian loss transformation derived above, the position quadrature variance of a momentum-squeezed vacuum state becomes

\begin{align}
\operatorname{Var}(q)_\eta = \eta\operatorname{Var}(q)
+ \frac{1-\eta}{2}.
\end{align}

Substituting the squeezed-state variance,

\begin{align}
\operatorname{Var}(q) = \frac12e^{2 \kappa},
\end{align}

gives

\begin{align}
\operatorname{Var}(q)_\eta = \frac12 \left(\eta e^{2 \kappa} + 1-\eta\right) \label{eq:sq_var_loss1}
\end{align}

Hence,

\begin{align}
F_Q^{(\eta)} \approx 4A^2 \left( \eta e^{2 \kappa} + 1-\eta \right). \label{eq:loss_sq_qfi}
\end{align}
 where $A$ is as in \eqref{eq:cases_A}.

As in the lossless case, the distinction between the SU(2) and SU(1,1) interferometers enters only through the scale factor $A$, while optical loss modifies the QFI solely through the transformed quadrature variance.

\section{Quantum Fisher Information in Terms of the Mean Photon Number}\label{app:QFI_meanphoton}

To compare the phase sensitivities of finite-energy GKP states and squeezed vacuum states at fixed energy, it is convenient to express the quantum Fisher information in terms of the mean photon number rather than the state parameters that define each input state. This appendix rewrites the lossless and approximate lossy QFI expressions derived in Appendices~C and D in terms of the mean photon number for both the input states.

\subsection{Mean Photon Number}

For any single-mode state with vanishing first moments $\langle q\rangle = \langle p\rangle = 0$ and the mean photon number is $\bar n = \langle a^\dagger a\rangle$. Using

\begin{align}
a = \frac{q+ip}{\sqrt2},\qquad a^\dagger = \frac{q-ip}{\sqrt2},
\end{align}

we obtain

\begin{align}
a^\dagger a &= \frac12 (q-ip)(q+ip),\\
&= \frac12 \left(q^2+p^2+i[q,p] \right).
\end{align}

Since $[q,p]=i$, it follows that
\begin{align}
{a^\dagger a = \frac12 \left(q^2+p^2-1\right).}
\end{align}

Taking the expectation value and using
$\langle q\rangle=\langle p\rangle=0$ gives

\begin{align}
{\bar n = \frac12 \left[ \operatorname{Var}(q) +
\operatorname{Var}(p) - 1\right].}
\label{eq:mean_photon_general}
\end{align}

Equation~(\ref{eq:mean_photon_general}) provides a general relation between the mean photon number and the quadrature variances for any dark port input with vanishing first moments. In the following, this expression is applied to finite-energy GKP states and squeezed vacuum states to rewrite the corresponding QFI expressions in terms of the mean photon number.

\subsection{Finite-Energy GKP States}


For the momentum-squeezed finite-energy GKP state considered in this work, the quadrature variance as is derived in Equation \eqref{eq:exact_unsqueezed_gkp_probability}. 
Since momentum squeezing transforms the quadrature operators
according to $\hat{S}^{\dagger} (\kappa)\hat{q}\hat{S}(\kappa) = e^{\kappa}\hat{q}$ and $\hat{S}^{\dagger}(\kappa)\hat{p}\hat{S}(\kappa) = e^{-\kappa}\hat{p}$, the momentum variance is obtained from the corresponding
unsqueezed variance by replacing $e^{\kappa}$ with $e^{-\kappa}$. Substituting these expressions into
(\ref{eq:mean_photon_general}) yields

\begin{align}
\bar n_{\rm GKP} &= \frac{1}{2} \left(e^{2\kappa} + e^{-2\kappa}\right) \mathrm{Var}(q) - \frac12, \\
&= 
\cosh(2\kappa) \mathrm{Var}(q) - \frac12.
\end{align}

Hence,

\begin{align}
\mathrm{Var}(q) = \frac{(\bar n_{\rm GKP}+1/2)}{\cosh(2\kappa)}.
\label{eq:delta_mean_photon}
\end{align}

\subsubsection{Quantum Fisher Information without Optical Loss}

The preceding result allows the finite-energy parameter to be eliminated in favour of the experimentally meaningful quantity $\bar n_{\rm GKP}$. Using the lossless QFI derived in Appendix~C,

\begin{align}
F_Q = 8A^2e^{2\kappa} \mathrm{Var}(q),
\end{align}

and substituting (\ref{eq:delta_mean_photon}), we obtain

\begin{align}
F_Q &= 8A^2 (\bar n_{\rm GKP}+1/2)\frac{e^{2\kappa}}{\cosh(2\kappa)}.
\end{align}

Using the identity

\begin{align}
\frac{e^{2\kappa}} {\cosh(2\kappa)} = \frac{2}{1+e^{-4\kappa}},
\end{align}

the quantum Fisher information can be written as

\begin{align}
{F_Q = 16A^2 \frac{\bar n_{\rm GKP}+1/2}{1+e^{-4\kappa}}.}
\label{eq:qfi_mean_photon}
\end{align}
Equation~(\ref{eq:qfi_mean_photon}) gives the lossless quantum Fisher information of a finite-energy GKP state entirely in terms of the mean photon number, the squeezing parameter, and the interferometer-dependent scale factor $A$.

\subsubsection{Approximate Quantum Fisher Information with Optical Loss
}

Using the approximate QFI under optical loss in Equation \eqref{eq:loss_gkp_qfi}, the quantum Fisher
information under optical loss is given by

\begin{align}
F_Q^{(\eta)} \approx 8A^2 \left[ \eta e^{2\kappa} \mathrm{Var}(q)
+ \frac{1-\eta}{2}\right].
\end{align}

Substituting Equation (\ref{eq:delta_mean_photon}) into the approximate lossy QFI expression eliminates the finite-energy parameter in favour of the mean photon number,

\begin{align}
\mathrm{Var}(q) = \frac{(\bar n_{\rm GKP}+1/2)} {\cosh(2\kappa)},
\end{align}

the lossy quantum Fisher information becomes

\begin{align}
F_Q^{(\eta)} &\approx 8A^2 \left[\eta \frac{e^{2\kappa}}
{\cosh(2\kappa)}(\bar n_{\rm GKP}+1/2) + \frac{1-\eta}{2}
\right].
\end{align}

Using the identity

\begin{align}
\frac{e^{2\kappa}} {\cosh(2\kappa)} = \frac{2}{1+e^{-4\kappa}},
\end{align}

we obtain

\begin{align}
{F_Q^{(\eta)} \approx 16A^2 \eta \frac{\bar n_{\rm GKP}+1/2}{1+e^{-4\kappa}} + 4A^2(1-\eta).}
\label{eq:loss_qfi_mean_photon}
\end{align}

Equation~(\ref{eq:loss_qfi_mean_photon}) gives the approximate quantum Fisher information under optical loss in terms of the mean photon number, enabling direct comparison with squeezed vacuum states at fixed input state energy.

\subsection{Finite-Energy GKP States with Independent Peak and
Envelope Widths}
\label{app:GKP_mean_photon_two_widths}

The preceding expressions assume the one-parameter finite-energy GKP
family, for which the peak and envelope widths are not independent.
We now generalize the analysis to the finite-energy GKP state
parameterized by independent peak and envelope widths,
\begin{align}
\braket{q}{\mathrm{GKP}_{\sigma_{\mathrm{peak}},
\sigma_{\mathrm{env}}}}
= &
\mathcal{N}_{\sigma_{\mathrm{peak}},\sigma_{\mathrm{env}}}
\exp\left( -\frac{q^2}{2\sigma_{\mathrm{env}}^2} \right) \nonumber \\
&\sum_{s=-\infty}^{\infty}
\exp\left[
-\frac{
\left(q-s\sqrt{2\pi}\right)^2
}{
2\sigma_{\mathrm{peak}}^2
}
\right].
\label{eq:gkp_two_widths_mean_photon_state}
\end{align}
Here, $\sigma_{\mathrm{peak}}$ determines the width of the individual
lattice peaks, while $\sigma_{\mathrm{env}}$ determines the width of
the global Gaussian envelope. The two parameters are treated as
independent.

In this more general parameterization, the position variance is given
by the exact expression derived in
Appendix~\ref{app:gkp_variance_two_parameters},
\begin{align}
\mathrm{Var}(q)
&=
\frac{
\sigma_{\mathrm{peak}}^2
\sigma_{\mathrm{env}}^2
}{
2\left(
\sigma_{\mathrm{peak}}^2+
\sigma_{\mathrm{env}}^2
\right)
}
\nonumber\\
&\quad+
\frac{
\pi\sigma_{\mathrm{env}}^4
}{
2\left(
\sigma_{\mathrm{peak}}^2+
\sigma_{\mathrm{env}}^2
\right)^2
}
\frac{
\displaystyle
\sum_{s,t=-\infty}^{\infty}
(s+t)^2 W_{st}
}{
\displaystyle
\sum_{s,t=-\infty}^{\infty}
W_{st}
},
\label{eq:gkp_vq_two_widths_mean_photon}
\end{align}
where
\begin{equation}
W_{st}
=
\exp\left[
-\frac{\pi(s-t)^2}
{2\sigma_{\mathrm{peak}}^2}
-\frac{\pi(s+t)^2}
{2\left(
\sigma_{\mathrm{peak}}^2+
\sigma_{\mathrm{env}}^2
\right)}
\right].
\label{eq:gkp_pair_weight_two_widths_mean_photon}
\end{equation}

Unlike the one-parameter GKP state, the independent variation of
$\sigma_{\mathrm{peak}}$ and $\sigma_{\mathrm{env}}$ does not in
general guarantee that the unsqueezed position and momentum variances
are equal. The momentum variance can be evaluated directly from the
position-space wavefunction according to
\begin{equation}
\mathrm{Var}(p)
=
\langle p^2\rangle
=
\int_{-\infty}^{\infty}
dq\,
\left|
\frac{d\psi(q)}{dq}
\right|^2,
\label{eq:gkp_vp_derivative_mean_photon}
\end{equation}
where $\langle p\rangle=0$ for the real and symmetric wavefunction in
Eq.~(\ref{eq:gkp_two_widths_mean_photon_state}). Carrying out the
Gaussian integrations gives
\begin{align}
\mathrm{Var}(p)
&=
\frac{
\sigma_{\mathrm{peak}}^2+
\sigma_{\mathrm{env}}^2
}{
2\sigma_{\mathrm{peak}}^2
\sigma_{\mathrm{env}}^2
}
\nonumber\\
&\quad-
\frac{
\pi
}{
2\sigma_{\mathrm{peak}}^4
}
\frac{
\displaystyle
\sum_{s,t=-\infty}^{\infty}
(s-t)^2 W_{st}
}{
\displaystyle
\sum_{s,t=-\infty}^{\infty}
W_{st}
}.
\label{eq:gkp_vp_two_widths_mean_photon}
\end{align}

The mean photon number of the unsqueezed generalized GKP state is
therefore
\begin{equation}
\bar n_{\mathrm{GKP}}
=
\frac12
\left[
\mathrm{Var}(q)
+
\mathrm{Var}(p)
-1
\right].
\label{eq:gkp_nbar_two_widths_unsqueezed}
\end{equation}

For the momentum-squeezed GKP state, the quadrature operators
transform according to
\begin{equation}
\hat{S}^{\dagger}(\kappa)\hat{q}\hat{S}(\kappa)
=
e^{\kappa}\hat{q},
\qquad
\hat{S}^{\dagger}(\kappa)\hat{p}\hat{S}(\kappa)
=
e^{-\kappa}\hat{p}.
\end{equation}
Consequently, the squeezed-state variances are
\begin{equation}
\mathrm{Var}_{\kappa}(q)
=
e^{2\kappa}\mathrm{Var}(q),
\qquad
\mathrm{Var}_{\kappa}(p)
=
e^{-2\kappa}\mathrm{Var}(p).
\end{equation}
Substitution into Eq.~(\ref{eq:mean_photon_general}) gives
\begin{align}
{
\bar n_{\mathrm{GKP}}
=
\frac12
\left[
e^{2\kappa}{\mathrm{Var}(q)}
+
e^{-2\kappa}{\mathrm{Var}(p)}
-1
\right].
}
\label{eq:gkp_nbar_two_widths}
\end{align}

Equation~(\ref{eq:gkp_nbar_two_widths}) is the appropriate
mean-photon-number relation when the peak and envelope widths are
varied independently. In general, the mean photon number depends on
both quadrature variances and therefore cannot be determined from
$\mathrm{Var}(q)$ alone.

\subsubsection{Quantum Fisher Information without Optical Loss}

For the displacement-sensing configuration considered here, the
lossless QFI of the momentum-squeezed GKP state is
\begin{equation}
F_Q^{\mathrm{GKP}}
=
8A^2\mathrm{Var}_{\kappa}(q)
=
8A^2e^{2\kappa}\mathrm{Var}(q).
\label{eq:gkp_qfi_two_widths}
\end{equation}

Equation~(\ref{eq:gkp_nbar_two_widths}) can be rearranged to express
the position variance in terms of the mean photon number and the
unsqueezed momentum variance:
\begin{align}
e^{2\kappa}\mathrm{Var}(q)
&=
2\bar n_{\mathrm{GKP}}
+1
-e^{-2\kappa}\mathrm{Var}(p).
\label{eq:gkp_vq_from_nbar_two_widths}
\end{align}
Consequently, the lossless QFI becomes
\begin{align}
{
F_Q^{\mathrm{GKP}} = 8A^2 \left[ 2\bar n_{\mathrm{GKP}}
+1 -e^{-2\kappa}\mathrm{Var}(p)\right]. } \label{eq:gkp_qfi_mean_photon_two_widths}
\end{align}

Thus, unlike the one-parameter GKP family, the QFI of the
independently parameterized state cannot in general be written solely
as a function of $\bar n_{\mathrm{GKP}}$ and $\kappa$. At fixed mean
photon number, different choices of
$(\sigma_{\mathrm{peak}},\sigma_{\mathrm{env}})$ can lead to
different values of $\mathrm{Var}(p)$ and hence different QFI. This
residual dependence reflects the fact that the two-width
parameterization allows the available energy to be distributed
differently between the two quadratures.

\subsubsection{Approximate Quantum Fisher Information with Optical Loss}

Using the approximate lossy QFI derived in Appendix~D,
\begin{equation}
F_Q^{(\eta),\mathrm{GKP}} \approx 8A^2
\left[ \eta e^{2\kappa}\mathrm{Var}(q) + \frac{1-\eta}{2}
\right], \label{eq:gkp_loss_qfi_two_widths}
\end{equation}
and substituting Eq.~(\ref{eq:gkp_vq_from_nbar_two_widths}), we obtain
\begin{align}
F_Q^{(\eta),\mathrm{GKP}} &\approx 8A^2  \left[ \eta
\left( 2\bar n_{\mathrm{GKP}} +1 -e^{-2\kappa}\mathrm{Var}(p) \right)
+ \frac{1-\eta}{2} \right].
\end{align}
Equivalently,
\begin{align}
{ F_Q^{(\eta),\mathrm{GKP}} \approx 8A^2 \left[
2\eta\bar n_{\mathrm{GKP}} + \frac{1+\eta}{2} - \eta  e^{-2\kappa}\mathrm{Var}(p)\right]. }
\label{eq:gkp_loss_qfi_mean_photon_two_widths}
\end{align}

The lossy QFI therefore retains the same dependence on the
unsqueezed momentum variance as the lossless result. Optical loss
reduces the contribution of the input state variance by the factor
$\eta$ and adds the vacuum contribution $(1-\eta)/2$ inside the
bracket. The dependence on $\mathrm{Var}(p)$ remains because the mean
photon number constrains the sum of the two squeezed quadrature
variances rather than the position variance alone.

\subsection{Squeezed Vacuum States}

For a momentum-squeezed vacuum state with squeezing parameter $\kappa$, the mean photon number is

\begin{align}
\bar n_{\rm sq} = \sinh^2 \kappa.
\end{align}

Using

\begin{align}
\cosh(2\kappa) = 2\bar n_{\rm sq}+1,
\end{align}

together with

\begin{align}
e^{2 \kappa} = \cosh(2 \kappa) + \sinh(2 \kappa),
\end{align}

and

\begin{align}
\sinh(2 \kappa) = 2\sinh \kappa \cosh \kappa
= 2\sqrt{\bar n_{\rm sq}(\bar n_{\rm sq}+1)},
\end{align}

we obtain

\begin{align}
{e^{2 \kappa} = 2\bar n_{\rm sq} + 1 + 2\sqrt{\bar n_{\rm sq}(\bar n_{\rm sq}+1)}.}
\label{eq:e2r_mean_photon}
\end{align}

\subsubsection{Quantum Fisher Information without Optical Loss}
Substituting this into Equation (\ref{eq:sq_qfi}) gives

\begin{align}
F_Q = 4A^2 \left[2\bar n_{\rm sq} + 1 + 2\sqrt{\bar n_{\rm sq}(\bar n_{\rm sq}+1)}\right].
\end{align}

Hence,

\begin{align}
{F_Q = 4A^2 \left[2\bar n_{\rm sq} + 1 + 2\sqrt{\bar n_{\rm sq}(\bar n_{\rm sq}+1)}\right].}
\label{eq:sq_qfi_mean_photon}
\end{align}

This expression provides the quantum Fisher information of a squeezed vacuum state solely in terms of the mean photon number and the interferometer-dependent scale factor. It therefore enables a direct comparison with the corresponding finite-energy GKP expressions at fixed
mean photon number.

\subsubsection{Approximate Quantum Fisher Information with Optical Loss}

Using the approximate QFI under optical loss derived in Appendix~D, the quantum Fisher information under optical loss is given by

\begin{align}
F_Q^{(\eta)} \approx 4A^2 \left( \eta e^{2 \kappa} + 1-\eta\right).
\end{align}

Substituting Equation (\ref{eq:e2r_mean_photon}),

\begin{align}
e^{2 \kappa} = 2\bar n_{\rm sq} + 1 + 2\sqrt{\bar n_{\rm sq} (\bar n_{\rm sq}+1)},
\end{align}

yields

\begin{align}
F_Q^{(\eta)} &\approx 4A^2 \left[ \eta \left(2\bar n_{\rm sq} + 1 + 2\sqrt{\bar n_{\rm sq} (\bar n_{\rm sq}+1)}\right) + 1-\eta \right].
\end{align}

Combining the constant terms gives

\begin{align}
{F_Q^{(\eta)} \approx 4A^2 \left[1 + 2\eta \left(\bar n_{\rm sq} + \sqrt{\bar n_{\rm sq} (\bar n_{\rm sq}+1)
}\right)\right].}
\label{eq:sq_loss_qfi_mean_photon}
\end{align}

Equation~(\ref{eq:sq_loss_qfi_mean_photon}) expresses the approximate
quantum Fisher information under optical loss solely in terms of the
mean photon number, the optical transmissivity, and the
interferometer-dependent scale factor.

\end{document}